# VESPA: a community-driven

# Virtual Observatory in Planetary Science

**S. Erard** (1), B. Cecconi (1), P. Le Sidaner (2), A. P. Rossi (3), M. T. Capria (4), B. Schmitt (5), V. Génot (6), N. André (6), A. C. Vandaele (7), M. Scherf (8), R. Hueso (9), A. Määttänen (10), W. Thuillot (11), B. Carry (11), N. Achilleos (12), C. Marmo (13), O. Santolik (14), K. Benson (12), P. Fernique (15), L. Beigbeder (16), E. Millour (17), B. Rousseau (1), F. Andrieu (1), C. Chauvin (2), M. Minin (3), S. Ivanoski (4), A. Longobardo (4), P. Bollard (5), D. Albert (5), M. Gangloff (6), N. Jourdane (6), M. Bouchemit (6), J.-M. Glorian (6), L. Trompet (7), T. Al-Ubaidi (8), J. Juaristi (9), J. Desmars (11), P. Guio (12), O. Delaa (13), A. Lagain (13), J. Soucek (14), D. Pisa (14)

(1) LESIA, Observatoire de Paris, PSL Research University, CNRS, Sorbonne Universités, UPMC Univ. Paris 6, Univ. Paris Diderot, Sorbonne Paris Cité, France (2) DIO-PADC, Observatoire de Paris, PSL Research University, France (3) Jacobs University, Bremen, Germany (4) IAPS, INAF, Rome, Italy (5) IPAG UGA/CNRS, Grenoble, France (6) IRAP, Université Paul Sabatier, CNRS, Toulouse, France (7) IASB/BIRA, Brussels, Belgium (8) IWF, OeAW, Graz, Austria (9) Dpto. Física Aplicada I, Escuela de Ingeniería de Bilbao, UPV/EHU, 48013, Bilbao, Spain (10) LATMOS/IPSL, UVSQ, Université Paris-Saclay, UPMC Univ. Paris 06, CNRS, Guyancourt, France, (11) IMCCE, Observatoire de Paris, PSL Research University, CNRS, France (12) University College London, U-K (13) GEOPS/CNRS/ U. Paris-Sud, Univ. Paris Saclay, Orsay, France (14) Institute of Atmospheric Physics, Czech Academy of Sciences, Prague, Czechia, (15) Observatoire astronomique de Strasbourg, Université de Strasbourg, UMR 7550, France, (16) GFI, Toulouse, France, (17) LMD / IPSL, CNRS, Paris, France

---

*Corresponding author:* Stéphane Erard

Full Address:  LESIA / Observatoire de Meudon
 5 pl Jules Janssen
 92195 Meudon, France

Tel: 33 1 45 07 78 19

E-mail: stephane.erard@obspm.fr

Keywords: Virtual Observatory; Solar System; GIS

## Abstract

The VESPA data access system focuses on applying Virtual Observatory (VO) standards and tools to Planetary Science. Building on a previous EC-funded Europlanet program, it has reached maturity during the first year of a new Europlanet 2020 program (started in 2015 for 4 years). The infrastructure has been upgraded to handle many fields of Solar System studies, with a focus both on users and data providers. This paper describes the broad lines of the current VESPA infrastructure as seen by a potential user, and provides examples of real use cases in several thematic areas. These use cases are also intended to identify hints for future developments and adaptations of VO tools to Planetary Science.

## Introduction

Modern observational programs in Planetary Science produce huge datasets, especially with long-lived space missions orbiting planets, ground-based campaigns and simulation codes. This calls for new ways to handle the data, not only to perform mass processing, but also more basically to access them easily and efficiently. Virtual Observatory (VO) techniques developed in Astronomy during the past 15 years can be adapted to address this problem, provided they are enlarged to support specificities of Solar System studies. These include body-related coordinate systems and time scales (local time and season), or measured quantities related to reflected light, but also to physical samples such as meteorites or planetary samples. An effort to adapt VO techniques to Solar System studies has been started in the frame of the EC-funded Europlanet program, first as a demonstrator in the 7[th] Frame Program (FP7, 2009-2012) (Erard et al., 2014a), and now in a more extensive way in the current Frame Program Horizon 2020 (2015-2019). In the on-going program Europlanet 2020, the VESPA activity deals with the infrastructure and implements new data services. *VESPA* stands for Virtual European Solar and Planetary Access, and supports all aspects of Solar System science. Its developments are essentially based on the standards of the *IVOA* (International Virtual Observatory Alliance), but also include elements from more specific areas, most notably *IPDA* (International Planetary Data Alliance, focusing on planetary archives from space agencies), *HELIO* (Heliophysics Integrated Observatory, focusing on Solar physics and Sun-Earth interactions), *IMPEx* (Integrated Medium for Planetary Exploration, focusing on planetary plasma physics), *SPASE* (Space Physics Search and Extract, proposing a unified registry for space plasma physics in the Solar System), and various efforts to adapt Geographic Information Systems (*GIS*) to Planetary Science, as well as standards issued by the Open Geospatial Consortium (*OGC*). VESPA is of course also responsive to standards maintained by the International Astronomical Union (*IAU*).

The paradigm inherited from the VO is to set up the infrastructure for a distributed, collaborative system to which any research group or institute can contribute by sharing their data or providing new tools, relying on internet connections and distributed local services, rather than big and expensive data centers concentrating data from many origins. The underlying principle is that the data remain where the science expertise is located, and the system connects services that are maintained as part of the standard web facilities existing in every institute. Building the VO is therefore also about developing efficient interfaces between data services and tools.

The baseline for VESPA activity is to use previous developments, make them compatible and working together, and only add whatever is missing to handle Solar System data efficiently. The first step was to develop an adequate data distribution system allowing for searches based on parameters that make sense for planetary scientists. This implies a Data Model to describe the data accurately, a way to track and locate existing data services, a method to query them and retrieve data of interest, and tools to display data and start processing them. The most important action of course is to put relevant and accurate data on line and make them accessible through the current system. This is fully open to independent contributions, according to basic VO principles.

A specificity of VESPA, compared to other VO-related environments, is that all services present a uniform interface and are by default queried together. This is intended to favor data discovery and to provide more visibility to smaller services, although other query modes allow the user to access specific descriptions in Planetary Science services, and to aggregate such services with other ones from the realm of Astronomy. Another unique capacity is to provide a link with GIS, which are more and more used for the study of high resolution imaging of planetary surfaces.

This paper first describes the broad lines of the current VESPA infrastructure as seen by a potential user. Section 2 describes the origin of current data services and strategies to include more data in this environment. Section 3 then provides examples of real use cases in each thematic area. This not only illustrates the capacities of this system, but also outlines the need for additional functionalities to be implemented in the near future. Those are summarized and discussed in Section 4.

To preserve legibility, standards and protocols are described in Annex 1 with links, while tools are listed in Annex 2 with references and links. The corresponding acronyms are introduced in the text with italics (standards) and bold face (tools) respectively. Existing VESPA-enabled data services are listed in Table 1 with references.

# 1- The VESPA environment

## 1.1 VESPA current environment

The VESPA data access architecture is based on a new data access protocol, a specific user interface to query the available services, and intensive usage of tools and standards developed for the Astronomy VO (Fig. 1). The Europlanet data access protocol, *EPN-TAP*, relies on the generic TAP mechanism associated to a set of parameters that describe the contents of a data service. *TAP* (Table Access Protocol) is an IVOA protocol to query tabular data such as catalogs, e. g., a list of files in a dataset. The EPN-TAP parameters (Erard et al., 2014b) are defined to enable queries on quantities relevant to the scientific user, including observational and instrumental conditions, and to handle the specific diversity and complexity of Planetary Science. Data services return answers formatted as *VOTable*, a self-described file format which is handled by all standard VO tools.

Data services are installed at their respective providers institutes and are declared in the standard IVOA registries. The registries provide catalogs of available services with their characteristics, including the access protocols in use (Demleitner et al., 2014), so that they are always visible and reachable from query interfaces. Although accessible in many ways, EPN-TAP data services are best queried from the optimized VESPA main user interface, or portal, maintained at Paris Observatory: http://vespa.obspm.fr

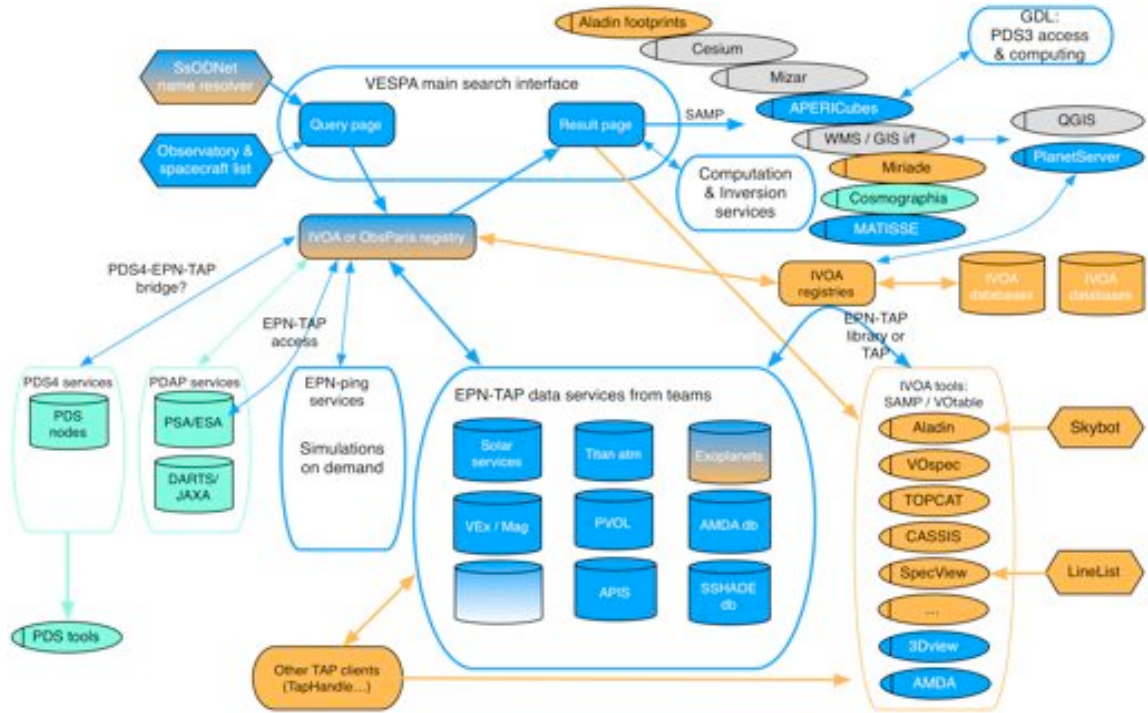

*Figure 1: short-term roadmap for the VESPA ecosystem. Colors indicate the main origin of the developments: orange = IVOA, blue = VESPA, light green = PDS / SPICE / IPDA, silver = OGC / GIS-related. Bold arrows are connections already implemented, thin ones are under study or limited in scope.*

The VESPA main user interface is functionally divided in two parts: a query system, and a presentation of results. Several web forms are available to support various types of queries, but most of them present a set of fields used to enter EPN-TAP parameters that restrain the query. When the query is submitted, the VESPA user interface forwards it to all EPN-TAP services declared in the IVOA registries, and gathers their results. According to the TAP protocol, these results consist in a subset of lines in the service table that match the query conditions (Fig. 2). Each line includes either a link to a data file or a set of physical properties stored as parameters. By default, the result page only displays a selection of parameters describing the first results; this selection can be modified using the interface buttons on the top-left side, and multi-page navigation is available at the bottom-right side of the table.

The result page is currently presenting query results for individual services separately, and will be updated in the coming months to mix results from several services and facilitate cross-examination of various datasets. From the result page, information can easily be browsed and sent to standard VO tools for display and processing; this includes data, general metadata and spatial footprints. The most useful VO tools in VESPA's context are listed in Annex 2 with reference and link.

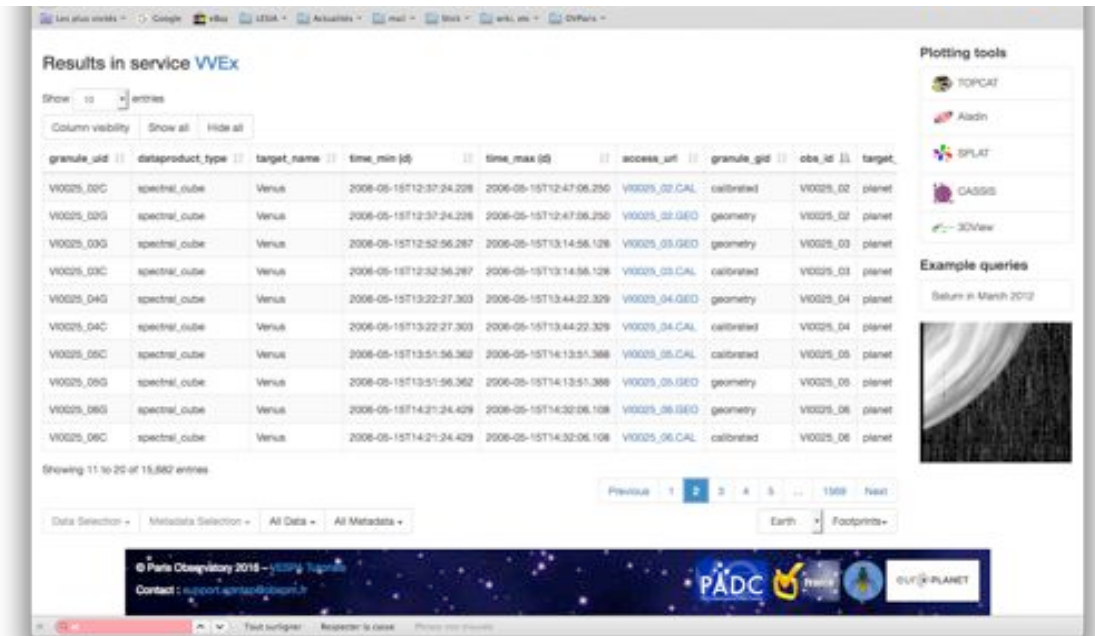

*Figure 2: VESPA result page from the VIRTIS/Venus-Express service, with thumbnails. Links to VO tools are provided on the right panel. Lower buttons (Data Selection…) provide tool connectivity.*

Several functions are implemented in the interface to make the search workflow more efficient and more friendly to occasional users. First of all, a Solar System objects name resolver (**SsODNet**) is used behind the query form to insure that a target will be searched under all its possible designations, which is particularly sensitive for small bodies; this approach is intended to address the problem of changing names and multiple designations in existing archives, and results in significant time saving and better completeness. In a similar way, data origin is documented by an Observatory and Spacecraft list built upon the IAU observatory list and other sources (Cecconi et al., 2015a). This list is currently being designed as an EPN-TAP service but it will also be used as a resolver, again to include all existing aliases in a query related to instruments or observing facilities.

A very practical concern with a modular system such as the VO is that the user has to be aware of the capacities of all available tools in order to use it properly, so that the learning curve may be very steep. This problem is addressed in two ways. First, the most relevant tools can be launched from the VESPA user interface; this relieves the users from the need to store these applications on their local system, and ensures that the latest version is downloaded when invoked (using a local instance is of course possible, and is a time saver for experienced users). Second, support applications displayed in Fig. 1 are launched automatically when a related function is used, as detailed in the next paragraph. Users therefore do not need to be familiar with the VO applications to use the search portal, and are only required to have a general knowledge of their functionalities.

Some actions in the result page trigger the launch of support applications: thumbnails (if provided in the service) are displayed on mouse-over in the page; whenever transferred, metadata are open in **TOPCAT** which provides very powerful functions to analyze and visualize tabular data; footprints on planetary surfaces (Fig. 3) are plotted in **Mizar**, **Cesium** or **Aladin** depending on format (bounding box or sampled contour); spectral cubes from the VIRTIS imaging spectrometer on Venus-Express and Rosetta are visualized in the **APERICubes** slicer (Fig. 4). Additional functions are being implemented in the result page of the VESPA portal, e. g., target position and phase at a given date will be retrieved thanks to the **Miriade** ephemeris generator. Altogether, the presence of support applications in the main interface significantly facilitates the use of the VESPA search system for occasional users.

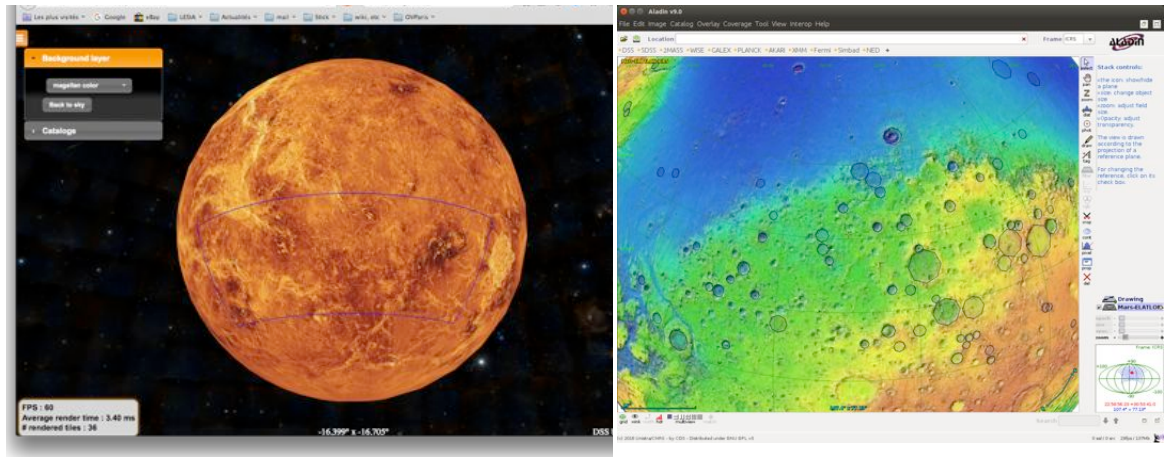

*Figure 3: left:a VIRTIS/Venus-Express cube footprint plotted in Mizar as bounding box (in purple); right: Mars crater footprints plotted in Aladin as sampled contours (in black)*

Detailed analyses of specific types of data can be performed with VO tools that need to be launched explicitly. Most tools implement the *SAMP* protocol from IVOA. *SAMP* (Simple Application Messaging Protocol) is a general mechanism to exchange data and messages seamlessly between software applications, including the VESPA search portal. Three existing tools are upgraded in the frame of Europlanet 2020 to better support Solar System data: **Aladin**, with implementation of planetary coordinate systems and a geo-extension to its standard sky coordinates format; **CASSIS**, with implementation of various types of reflectance spectra; **3Dview**, implementing *SPICE* kernels for all available space missions and adding mapping functions on planets.

In addition, interoperability is being improved in other cases: 3Dview can now receive specific VOTable (times series with a scalar or a 3-element vector quantity) to plot data along a spacecraft trajectory (Génot et al., 2017a); the tools included in the **AMDA** environment are now available to all users to plot time series retrieved from EPN-TAP data services; **MATISSE** will also receive a SAMP interface, to be used as 3D plotting tool on small bodies shape models; a GIS interface will also make it possible to retrieve planetary maps from WMS servers and send them to VO tools, or conversely to ingest data from VO services into GIS software, in particular **QGIS** and **PlanetServer** (see section 3 below). In a lesser interactive context, **Cosmographia** can be used to visualize trajectories provided as SPICE kernels by some services, e. g., the DynAstVO computation system for Near-Earth Objects in Paris (Desmars et al., 2016). As part of the training activity of VESPA in Europlanet 2020, tutorials are produced to illustrate key functions of the main tools in a Planetary Science context, and with data services related to several science fields. These tutorials are available from the VESPA web site (http://www.europlanet-vespa.eu). Besides, support to users is provided through a helpdesk (support.vespa@obspm.fr).

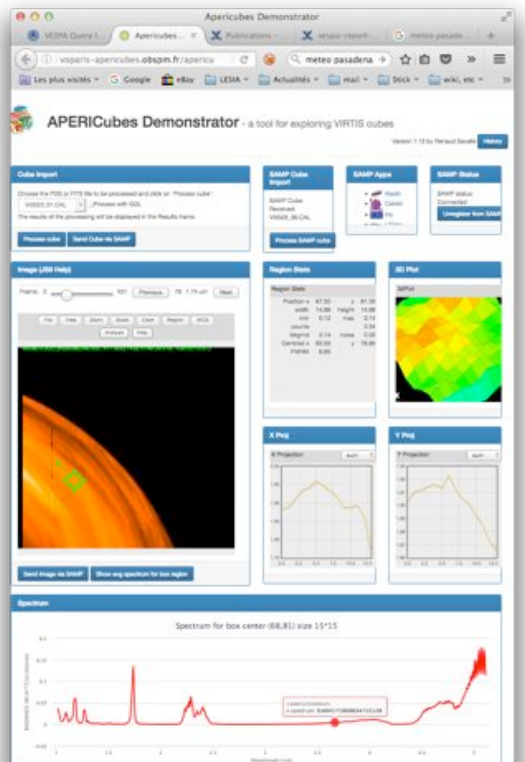

*Figure 4: VIRTIS/Venus-Express spectral cube of the night side sliced on-line in APERICubes. Spectra averaged in regions of interest (displayed on the bottom inlay) can be sent to CASSIS via SAMP.*

### 1.2 Alternative data access

Although optimized for global queries and service discovery, the VESPA main user interface is not the only point to access data. First of all, Planetary Science data services are TAP compliant and are therefore accessible individually from the usual TAP interface included in standard VO tools, and from dedicated search interfaces such as **TapHandle** (Fig. 9) or **TAP shell**. In addition, EPN-TAP libraries have been developed to query the data services directly from the VO tools, and have already been implemented in AMDA, 3Dview, and CASSIS. The **Propagation Tool** at IRAP/CDPP includes a more limited client to access some relevant services only, in particular APIS (Auroral Planetary Imaging and Spectroscopy, containing mostly HST observations of the giant planets). Finally, extra access modes have also been developed, such as a **Google Sheets add-on**, or basic python and **IDL/GDL** scripts, still to be completed. This multiplicity of accesses ensures that EPN-TAP services will remain reachable even if the main user interface becomes unavailable for some reason.

### 1.3 Short-term evolutions

A potential difficulty to extend the use of VO tools to Solar System data is related to the different data formats used in Planetary Science and Astronomy. All IVOA tools understand *VOTable*s and most variations of *FITS* (Flexible Image Transfer System), which is, among other things, the standard format for telescopic images. Some tools can ingest or produce other formats, e. g., *CDF* (Common Data Format, commonly used in plasma physics archives), plain ascii, CSV (Coma Separated Values), etc. However the usual space archive formats (PDS3 and PDS4) are not currently recognized, nor those used by GIS software. Supporting these data formats is therefore an obvious goal to enlarge VESPA capacities in the near future, and access to PDS format in particular is essential to ensure support of a large community and long-term sustainability. A PDS3 to FITS converter is already included in APERICubes, the usage of which will be enlarged in the future.

Uses cases presented in section 3 demonstrate that additional "computation services" need to be implemented to process search results. The most obvious example involves on-the-fly averaging of selected spectra or profiles, but more sophisticated services may include band identification in spectra, inversions of light curves and computation of Hapke photometric parameters, image segmentation and pattern extraction, etc. More complex processing pipelines can be implemented as workflows chaining several operations (Castelli et al., 2015), starting with data retrieval from one or various services. The use of workflows will make the reprocessing of entire datasets relatively easy.

Finally, powerful computing environments such as IDL or Matlab, widely used in the Planetary Science community, can be interfaced to a limited but useful extent by providing them with VOTable input/output capacities. VOtable input and output have been implemented for IDL (and its open source clone GDL) on Unix systems, essentially by scripting TOPCAT functions (alternative ways exist, in particular using the older **SSW** IDL library, but they are not fully operational in practice). This connection makes it possible to issue a query in the VESPA search interface and to pass the list of matching files to IDL for further processing. Conversely, tables prepared under IDL and stored as VOTable can easily be sent to VO tools - Aladin and TOPCAT in particular include very powerful display functions that are hardly reproduced in IDL (e. g., Fig. 13). Smooth interactions between usual computing environments and VO tools provide access to powerful processing methods, e. g., multivariate analyses on spectral cubes performed in GDL with results displayed in the more flexible VO tools. In specific cases, an IDL or GDL session can be launched on the server from the VO tools to perform dedicated processing transparently for the user (e. g., APERICubes internally converts from PDS3 to FITS format under IDL).

VESPA developments realized during the Europlanet program are released openly on our GitHub repository (https://github.com/epn-vespa).

# 2- Data content

## 2.1 Data services

During the first year of the Europlanet 2020 program, VESPA focused on implementing data services, with the goal to make more content available through interoperable access and have enough material to identify possible improvements related to data description. Since most datasets consists in lists of files or catalogs of observed properties, the preferred access protocol is EPN-TAP, which allows queries based on a set of parameters defining ranges on several axes (time, location, frequency, illumination conditions), target properties, measurement type, and various data identifiers to support callback mechanisms. The EPN-TAP protocol has been upgraded from version 1 (Erard et al., 2014b) to version 2 in 2016 (Erard et al, in preparation) to add flexibility and to define an extension procedure that will allow implementing new science fields in the future. Besides, credits have been cautiously taken care of, and are provided not only for global services, but also at the level of individual files. Several levels of attributions are given for service providers, publishers, and original data collectors.

New EPN-TAP data services are set up in various ways in Europlanet 2020: they are designed either by VESPA partners, by external teams selected after a public call for projects which is open every year in December, or by external contributors who wish to take advantage of VESPA data description and search functions. In addition, EC-funded programs producing Planetary Science related data are solicited to make them available through EPN-TAP services. This starts with the Trans-National Access activities in Europlanet 2020, which propose access to laboratory experiments and planetary analogue field on the basis of open calls.

Two large amateur databases have also been pre-identified to be provided with an EPN-TAP interface: PVOL (planetary images) and RadioJOVE (radio measurements of Jupiter). PVOL (Hueso et al., 2017) has been redesigned and upgraded from a pre-existing web site, and is now publicly open in Bilbao. RadioJOVE is a NASA-supported project started in 1998 (Higgins et al., 2014), with world-wide contributions; the EPN-TAP service has been set up in Paris and the database is now in place (Cecconi et al., 2015b). Both databases include a submission system from registered contributors and provide detailed references to the original observers. To help document the origin of data, the Observatory list will be open for amateurs to register and describe their facilities, as an update of a previous Europlanet FP7 project (**Matrix of ground-based facilities**, Scherf et al. (2011)). Other similar contributing services could be added in the frame of Europlanet to favor public outreach and participative science, in particular for the observation of comets and for large collections of planetary images. In fact, such amateur services nicely complement the historical professional collections already implemented or in project, e. g. a service of images collected during an IAU program in the 60's (BDIP, Drossart et al. (2002)).

### 2.2 Observational archives

Although the main focus of VESPA is on derived data, accessing archives of calibrated data (and in some cases even raw data) from space agencies and large telescope organizations has many benefits: first because the EPN-TAP interface provides powerful search functions to the archives themselves, which are seldom searchable in details at present as far as Solar System data are concerned; second, because complete datasets can profit from mass processing based on workflows, on condition that the operational and observational parameters are easily accessible.

In addition to EPN-TAP, the VESPA search interface supports the *PDAP* (Planetary Data Access Protocol) protocol of *IPDA*, which is currently implemented on ESA's Planetary Science Archive (*PSA*) and JAXA's Data Archive and Transmission System (*DARTS*). However, due to the current shortcomings in PDAP definitions and usage, this interface is not very efficient and only allows for basic searches.

In the short-term, the on-going installation of an EPN-TAP data service on the PSA (Martinez et al., 2016; Macfarlane et al., 2017) will make all European Solar System space archives readily searchable via VESPA. This will in particular allow for cross searches between several instruments on the same mission (e. g., simultaneous images and spectra), between similar types of observations from several spacecraft (e. g. images of a given region of Mars through time), and between space borne and ground-based observations, or experimental measurements (e. g., mineral spectroscopy) (Fig. 5). The PSA EPN-TAP interface is expected to become available in 2017 (Besse et al., 2017), and the assessment phase has triggered an intense activity to check the content of PSA datasets and make it more consistent.

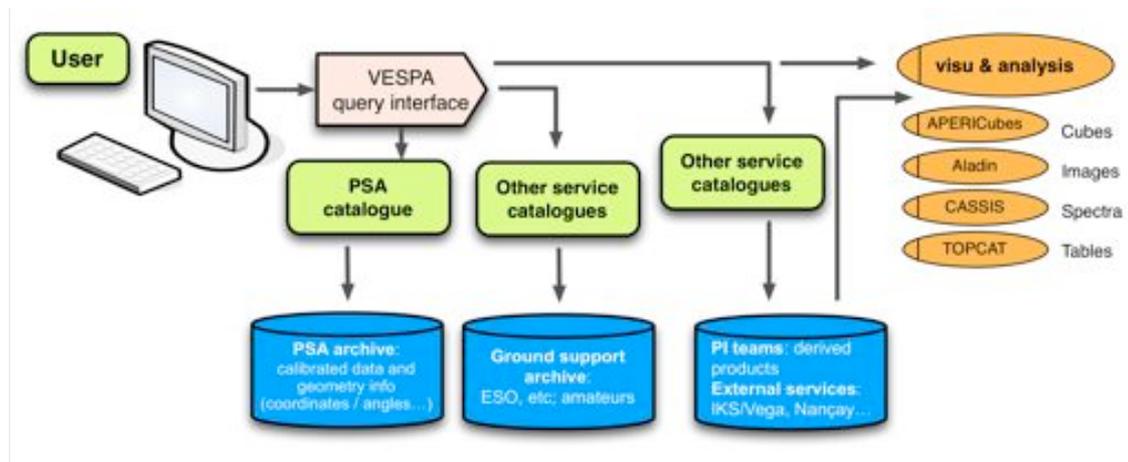

*Figure 5: EPN-TAP access to PSA datasets will be readily completed with other related data, including other space instruments and missions, ground-based support, and derived products.*

Several data services installed in research institutes were set up to validate the concept and link to PDS3 data files located either in the *PSA* (VIRTIS/Venus-Express), in the *PDS* Small Bodies Node (IKS/Vega-1), or in *DARTS* (datasets from the Selene mission, in test). In the future, reaching archives provided in PDS4 standard (e. g., from PDS nodes and new ESA missions) is another major topic. This will require either the installation of an EPN-TAP service on the archives, or the setup of a global EPN-TAP to PDS4 metadata query converter. Several assessment studies were conducted in this direction with some PDS nodes. Again, a major benefit would be to easily link space observations with ground-based, experimental, and simulation data.

**2.2.1 Enhancing archives with PI teams contribution**

Even with a global EPN-TAP interface available on the public archives, independent contributions from the teams will still be valuable in some cases, due to the granularity of the PDS datasets. In the example of VIRTIS/Venus-Express, the dataset provides a global description at the level of spectral cubes, which correspond to observing sessions and include some $10^4$ spectra each; depending on operational parameters, they can encompass large areas and mix a variety of geometrical configurations (local time, latitudinal coverage, altitude above the limb, etc). A project to enhance this data service consists in providing a description of individual spectra using their associated geometry parameters, which becomes essential for detailed interpretation of spectral observations of surfaces in particular (e.g. by VIRTIS/Rosetta, VIR/DAWN, VIMS/Cassini, Mars imaging spectrometers, etc). Although the information exists in the PDS dataset, it is very difficult to extract at the level of the archive itself and this is best implemented by the PI teams.

EPN-TAP also offers a nice and light framework to address the growing concern to make derived data from space missions widely available, To go on with the example of VIRTIS/Venus-Express, the aim would be to distribute maps of winds, profiles of molecular abundances, surface temperature maps, etc, so as to make them available to the community in digitized form after publication. This also provides the opportunity to share all derived products generated during such studies, beyond the restricted selection usually published in research papers. In fact, paper-related datasets can be distributed as EPN-TAP services, the same way as *VizieR* distributes astronomical catalogs, and more generally data attached to publications, in the IVOA.

### 2.3 Simulation services

Getting results of simulation codes in the same environment as observations is an important matter for the scientific user. For this reason, the VESPA team is now studying another simple protocol to call simulations on demand – the working name for this project is EPN-ping. In this case the TAP mechanism is not adapted, because the arguments are used as input parameters to a code rather than search parameters in a table, and can take any continuous value. The baseline of this protocol is to provide input arguments through the same parameters used by EPN-TAP, which must be interpreted and converted on the server side; the calling mechanism itself can be derived either from existing IMPEx developments, or from the *HAPI* protocol for time series described in section 3 (see Magnetospheres subsection). The service answer is a VOTable formatted similarly to an EPN-TAP query result, so that it can be handled just like a data service result, making comparisons very easy. The definition use case for this new protocol is the Mars Climate Database / atmospheric profiles example discussed in details in section 3.

### 2.4 External contributions, compatibility

The VESPA infrastructure has been designed to remain as open as possible. Any research team can benefit from the VESPA system to distribute their data in the VO, provided that the data service follows the EPN-TAP protocol and is properly declared in the IVOA registry of services; it will be immediately searchable through the main interface and with other access methods. Documents are available on the VESPA wiki site to describe the preferred implementation procedure (https://voparis-confluence.obspm.fr).

If we consider the core VESPA infrastructure to consist in EPN–TAP data services and VO tools (plus EPN-ping services in the future), external contributions may consist in existing IVOA data services, non-VO data services, and computation services.

By construction, astronomical IVOA data services are intrinsically interoperable with VESPA through the VO tools, although they do not benefit from the uniform interface provided by EPN-TAP. For instance, stellar spectra can be searched from TOPCAT or CASSIS with *SSA* (the Simple Spectral Access Protocol of IVOA), and then combined with data products retrieved from VESPA data or simulation services. This ability makes all astronomical VO data accessible for Solar System studies, with obvious applications to calibration pipelines for telescopic observations. Conversely, this VO interoperability will give Planetary Science data a larger visibility in the Astronomy world. This also extends to other environments and databases, e. g., the *LineList* protocol available in **Specview** provides access to libraries of atomic and molecular transitions from *VAMDC* related databases, which can be, e. g., plotted over UV auroral spectra from the APIS service (Lamy et al., 2015).

Non-VO data services are best connected to the VESPA ecosystem by providing them with an EPN-TAP interface, which is indeed a significant activity in this phase of expansion. This of course does not affect the availability of previous or alternative query systems, but provides the services with the extra capacity to be queried together with other EPN-TAP services in the same field. For instance the Encyclopedia of Extrasolar Planets in Paris (Schneider et al., 2011) also has its own web search interface and plotting tools, and is also searchable through the *Cone Search* protocol (based on sky coordinates) via another VO interface.

A lighter interfacing method is to restrain to compatibility through tools, typically by implementing SAMP (i. e., the capacity to push data to VO tools) and a standard VO output format, preferably VOTable or FITS in most cases. This solution is also applicable to external computation services, especially when their input interface is too sophisticated to be simply handled through EPN-ping. For instance, a web-based spectral simulation service issuing spectra formatted as VOTable, including units and axes described with the relevant *UTypes* and Unified Content Descriptors (*UCD*), can easily feed VO tools such as CASSIS or TOPCAT for comparison with data retrieved from EPN-TAP services, and would benefit from CASSIS spectral functions (modeling, resampling, etc). Adding SAMP connectivity to such services makes this process even more transparent, although they will not accept direct queries from the VESPA interface. A side action in VESPA is therefore to encourage managers of such on-line simulation services to output their data in a VO-compatible format.

## 3- Science themes and use cases

The VESPA activity in Europlanet 2020 is structured around science themes to favor implementation of relevant data services, to study larger services associated to computing systems, and to identify key analysis functions to be provided by tools in each field. Data services available as of writing are listed in Table 1 and grouped by science theme.

With several data services now available in each science theme, it becomes possible to implement use cases reproducing actual scientific studies, so as to identify missing functions and to complete the tools and procedures available. Example use cases are included in this section; some specific ones are presented more at length when they result in important requirements on the services or infrastructure; other use cases are discussed in the companion papers in this issue and are not repeated here, e. g., Génot et al. (2017a), Hueso et al. (2017), Cecconi et al. (2017), Trompet et al. (2017).

### 3.1 Small bodies

The small body theme currently includes several observational services encompassing multiple targets (M4ast, TNOsarecool, BaseCom) and two services of dynamical parameters (a VO version of the Minor Planets Center catalog and DynAstVO). So far, only one dataset dedicated to a single object is available (IKS/Vega-1), but it will soon be completed with Dawn- and Rosetta-related services. Other existing services will be connected through EPN-TAP in the coming months.

A hard point with this science theme is related to target names, which consist either in names, numbers, or provisional and possibly secondary designations. The problem is that none of these parameters (name, number, designations) is defined for every target, so that no single ID is currently available to identify accurately all targets in the Solar System. Although the SSODNET resolver handles multiple designations, providing a uniform naming scheme is currently an issue. Other problems are related to syntactic variations in designations with time, which are not currently standardized and suffer local variations in some services (space or underscore included in designations, plus variations since the beginning of the 20[th] century). Such problems have to be handled by IAU, but pending a global decision, alternative solutions include the current Minor Planet Center naming scheme (name, or provisional designation if no name) or Gaia mission's catalog ID. Using a clear naming scheme is of course pivotal in comparing data from several services, e. g., to select targets from their orbital parameters in one service, and to retrieve their spectra or photometric properties from another one.

An EPN-TAP extension was defined to describe orbital parameters consistently between services, as well as rotational parameters, disk attitude, dynamical type and spectral class. This extension, together with a consistent naming scheme, will make it possible to work smoothly with several services describing some $10^5$ targets. The Minor Planets Center catalog has been implemented as an EPN-TAP service, and provides orbital parameters updated daily from the original MPC site. DynAstVO is a different computing system currently restrained to NEOs, and also providing trajectories as SPICE kernels. The VESPA user interface will soon be linked to the Miriade VO service to retrieve position and physical ephemeris at the date of interest. Physical ephemeris will be coupled to a collection of 3D shape models to provide realistic visualization of these bodies, with the scope of preparing or analyzing space missions. MATISSE can also be used to display accurate mapping of resolved data on the 3D shapes, in particular in the case of space observations, while trajectories can be visualized with Cosmographia (Fig. 6).

Finally, a planning system will be setup for multisite observations of occultations by small bodies. This will use the Observatory list mentioned above, with associated visibility charts (minimum visible altitude as a function of azimuth) and ephemeris system. This service will generalize the VISION functionality of the Miriade service, which currently provides visibility charts from a single observatory.

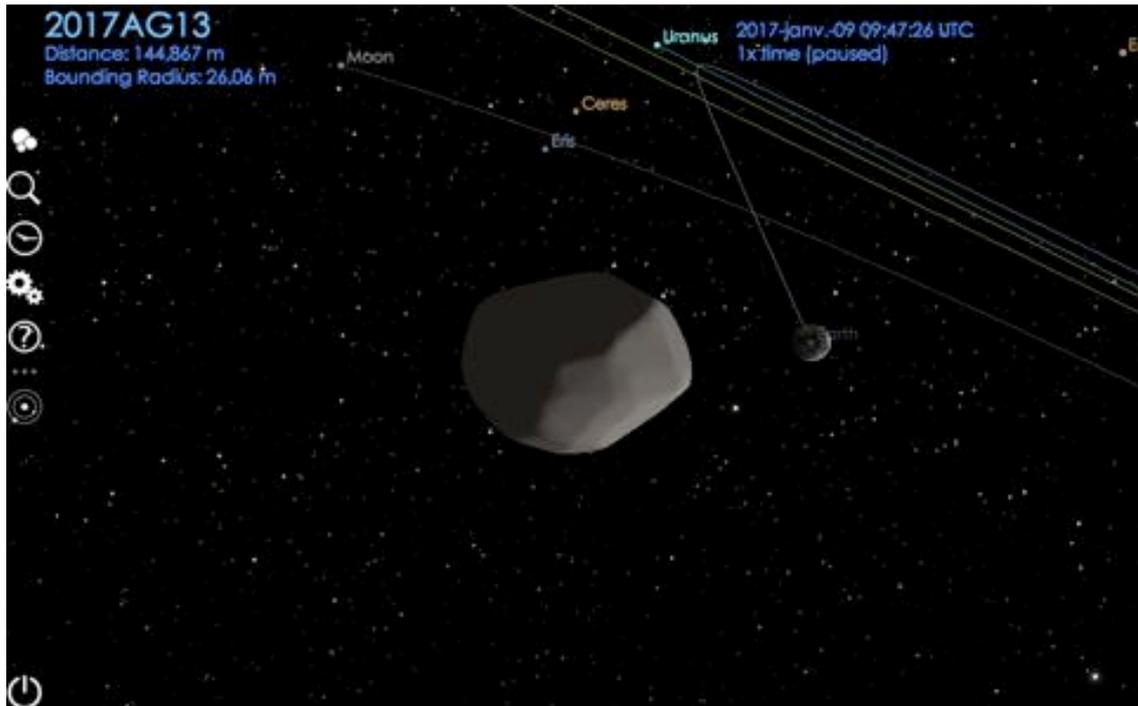

*Figure 6: view of asteroid 2017AG13 at closest approach to Earth well inside the Moon's orbit, as computed by DynAstVO and plotted in Cosmagraphia.*

### 3.1.1 Use case: spectral types of asteroid families

The M4ast service provides observational spectra of asteroids, but does not contain any information about spectral or dynamic class. Extracting spectra of a given class therefore relies on cross matches with other services containing such information, either MPC or DynAstVO. DynAstVO in particular uses a non-standard parameter "dynamical type" to specify NEO families. Such non-standard parameters can be accessed using the advanced query mode of the VESPA main interface. This mode is used here to select dynamical type strings containing "ATEN" in DynAstVO, which returns more than 1200 objects. The result table is uploaded to TOPCAT, together with all unique results from M4ast (formatted as VOtable only). The "Pair Match" function of TOPCAT is then used to cross-correlate the two tables on the basis of target names, which provides a new table containing references to spectra of Aten-class asteroids in M4ast; from there the spectra can be sent to CASSIS for plotting. Similar matches are possible for a spectral class (not currently available) or a range of orbital parameters, or to compare magnitudes provided in MPC and DynAstVO services for a class of NEOs.

However, although CASSIS can resample spectra to a common wavelength vector and perform basic arithmetic on pairs of spectra, no tool is currently available to average or combine many spectra in the general case; a solution is therefore being designed to compute the average spectrum from a list of objects, and to look for outliers.

## 3.2 Atmospheres

Most services currently installed in this field are related to vertical profiles of temperature and abundance, either measured (on Titan, Venus, and Mars) or simulated (extracts from the Mars Climate Database, MCD). In addition, the VIRTIS/Venus-Express calibrated dataset is searchable though EPN-TAP, and additional Mars-Express and Cassini derived data are currently in assessment phase. A service providing absorption cross-sections of various molecular species in the UV-visible range is being populated. Other extracts from the MCD could include daily curves at surface locations in the future.

Possible supporting tools include radiative transfer codes or inversion procedures applied to entire data services. It can be noted that 3Dview has the capacity to plot measurements in planetary atmospheres with geometric context, which very few tools are capable of. The vector plotting functions, although designed for magnetic fields, can be used to plot for instance a retrieved column density integrated along the viewing direction. This is potentially very helpful to visualize measurements in 3D configurations, such as a scan in 67P coma, or limb observations at Mars or Venus.

### 3.2.1 Use case: Mars Climate Database vs SPICAM profiles

The comparison between MCD and SPICAM profiles was an important use case to assess the comparisons between observations and simulations in a VO context. The MCD profiles were first installed as an EPN-TAP service, providing a selection of configurations sampled on a regular grid of coordinates, local time and season, for some of the available scenarios (a total of 62,244 profiles in this demonstration step). While the MCD code itself supports quasi-continuous variations in input parameters, sampling is required to implement data access via the TAP mechanism, which is originally designed to provide access to a catalog. Although the resolution of the service cannot be maintained in these conditions, a trade-off has been searched between the physical size of the resulting data service and the accuracy of the results.

The profiles themselves are provided as VOTable, a handy format for this type of data that is directly interpreted by VO tools. Parameters included in MCD profiles are temperature, pressure, densities of $H_2O$, $O_3$, and $CO_2$ computed in 50 layers up to 250 km altitude, plus column densities of water vapor and water ice, and dust opacity. SPICAM profiles include 3 different sets of $CO_2$ and temperature profiles in the 60-140 km range, derived under 3 different assumptions on the temperature at the top of the atmosphere (Forget et al., 2009). SPICAM measurements of $O_3$ are available separately in the 20-70 km range (Lebonnois et al., 2006). The service is being completed with aerosols profiles (Määttänen et al., 2013). The vertical scale is provided as radial distance and as altitude counted above both the ellipsoid and the actual surface (provided by MOLA topography) for the two services – this solution accounts for all possible scaling issue and will be extended to all similar atmospheric services.

The use case consisted in reproducing the study of SPICAM published in Forget et al. (2009) and evaluating if the current VO system makes it possible to perform similar comparisons with the MCD. Individual SPICAM profiles are readily selected in the VESPA user interface and sent to TOPCAT together with MCD results in the closest configuration (Fig. 7). Although this comparison validates the content of the services and illustrates the rapidity of comparing related data in a VO context, our fits are significantly poorer and probably less informative than in the original study. It appears that a natural workflow should include not only rejection of poor data, which is easily performed with the VO tools, but also monthly averages in extended areas to smooth out secondary variations; this is required to reproduce the much better match found by Forget et al. (2009) (their Fig. 16f). Similarly to the spectra handling described above, this stresses the need to develop an additional application to average the retrieved profiles, which is currently difficult in TOPCAT.

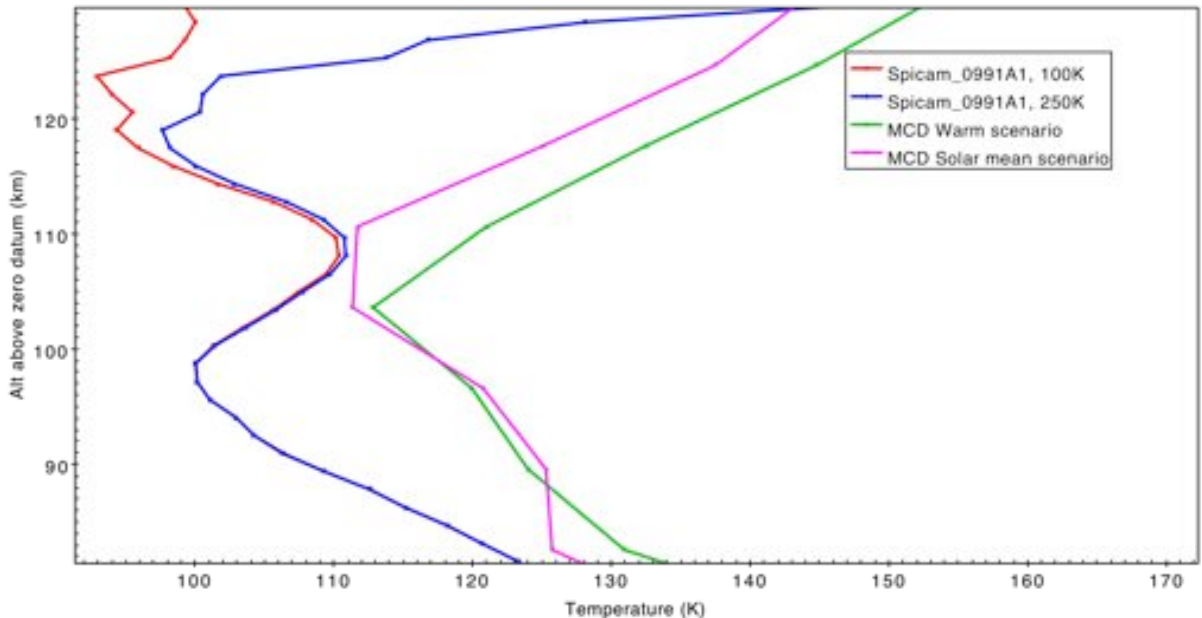

*Figure 7: SPICAM profile compared with MCD simulations in TOPCAT: SPICAM profiles are retrieved under different assumptions on temperature at the top of the atmosphere, MCD profiles are computed under two different solar flux scenarios (lon=348.4°, lat=44.4°, Ls=106.5°, local time=3.8h)*

Beyond the usual added value of VO access to observational data, this use case demonstrates the potential of comparing observations and simulations entirely in a VO environment. However, access to complex simulations via a strict TAP procedure appears to be restrictive. A naive approach requires sampling the parameters space and precomputing files corresponding to these situations. If the simulation uses many parameters, this is rapidly becoming inadequate: storing the files requires a large disk space but most of all, the details of the simulation are lost in the sampling procedure. For example, Fig. 8 shows a dramatic difference between MCD profiles extracted at the closest sampled point of the EPN-TAP service and in the exact configuration of SPICAM, which fits the measurements better. Although the MCD sampling step is much too coarse to preserve the accuracy of the simulation, this approach would already result in a significant data volume (~ 100 GB of precomputed files to accommodate all scenarios at the current resolution). Increasing the resolution of the service is required to meet the needs of scientific users, say by a factor of 10 on each of the four parameters, but this would lead to impracticable data volumes and still remain an approximation.

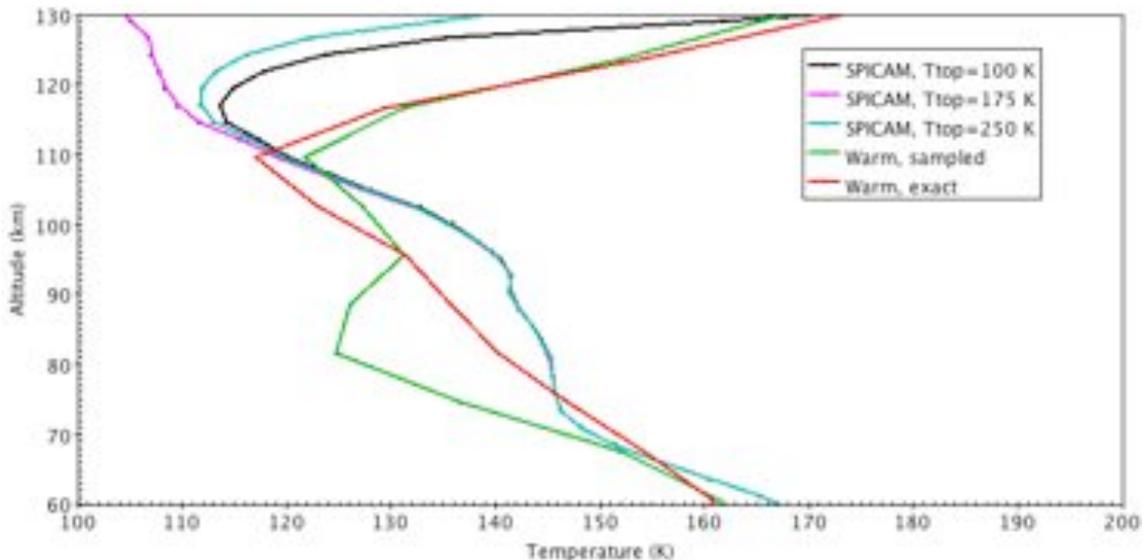

*Figure 8: SPICAM profiles compared with MCD simulations in TOPCAT: with exact parameters, or from closest sample. The sampled configuration is much further from the observations (lon=247.8°, lat= 44.4°, Ls=13.2°, local time= 21.7h)*

For this reason, the current EPN-TAP access to the MCD has been implemented differently. Instead of providing a link to a data file, like other EPN-TAP services do, this service calls a script that triggers on-line computation with the sampled parameters provided in the query, similarly to the standard on-line interface of the MCD. The resulting profile is formatted as a VOTable, which can be sent to TOPCAT for plotting. This script system eliminates the need to store pre-computed files and therefore allows for much smaller sampling steps, at the expense of computation on the server side. Although the EPN-TAP service only provides sampled configurations, the script can be called "manually" (i. e., in database language) from TOPCAT with any parameter values - in this case the exact SPICAM configuration (red curve in Fig. 8).

However, the need to preserve the flexibility of TAP access while maintaining the full resolution of the simulation has led to the study of another protocol to access simulation codes and computation-on-demand – the EPN-ping project already mentioned. The VESPA user interface will issue a direct call to a computing service, providing input arguments through the usual EPN-TAP parameters; on the server side, the EPN-ping interface will simply convert the input EPN-TAP parameters in code input parameters, and will format the output as a VOTable. Although similar to the current sampled TAP access, such a system would accept continuous values of the input parameters (like the standard MCD interface does). Additional functions could also be implemented more easily, e. g., performing averages when ranges of parameters are provided. This protocol will be studied in more details in the coming year. Possible applications include inversion codes, but go far beyond the single atmosphere theme.

### 3.3 Surfaces

Planetary surfaces are targeted by a very large set of experiments, many of which produce images and spectral cubes. Maps are routinely derived from mosaicked images, or from very large sets of single point measurements such as those produced by surface altimeters. Sparse measurements such as those performed by point spectrometers also provide important inputs for analyses of regional or global surface features.

Surface raster data can be distributed either as map-projected files in archives such as PDS and PSA, or alternatively as OGC web services (using protocols such as *WMS* for maps, *WCS* for coverages, etc); vector data such as toponyms, crater outlines or alike are also distributed in this framework (using, e. g., the *WFS* protocol, Rossi et al. (2016)). Such products are typically ingested in GIS tools such as **QGIS** or **PlanetServer** (Oosthoek et al., 2014; Marco Figuera et al., 2017). A goal of the surface theme in VESPA is therefore to bridge the standards and practices of Geographic Information Systems and the Virtual Observatory. It is involved in the very active *OpenPlanetary* initiative (Manaud et al., 2016b), which derives from ESA's first Planetary GIS workshop (Manaud et al., 2016a) and continued during the second (US) Planetary Data workshop (2015).

In order to make such resources discoverable and usable through the VO, using either VO or GIS/WebGIS tools, selected OGC web services have been connected to VESPA with an EPN-TAP interface. A first assessment has been performed with the planetary map server maintained by USGS Astrogeology in Flagstaff (Hare et al., 2017) (Fig. 9). In this case, the data products are global or regional maps built from many individual spacecraft images. The particularity of this EPN-TAP service is to provide a WMS interface instead of the usual link to a data product. Another assessment was performed with the MRO/CRISM service, which distributes individual calibrated image cubes (similar to the VIRTIS/Venus-Express service), again with a WMS interface instead of the URL of a data file.

*Figure 9: EPN-TAP table of the USGS planetary maps service, as seen in TapHandle.*

In both cases, the products can be searched from the VESPA user interface with the EPN-TAP parameters, just like any other service, but data access is different. Existing OGC web services made discoverable though VESPA are not suitable for standard VO tools, but instead for OGC-compliant services or desktop tools. Desktop plugins for Open Source Desktop GIS (e. g., QGIS) are being developed, and can connect GIS tools to the SAMP hub shared by the VO tools. This makes it possible to search an EPN-TAP service with a TAP interface such as VESPA or TapHandle, and send the results to QGIS for plotting the same way as with any VO tool (Fig. 10). A benefit of the EPN-TAP interface for surface studies is related to the very powerful search functions associated to IVOA-style footprints, which permits to look for intersections between complex polygonal shapes in spherical geometry.

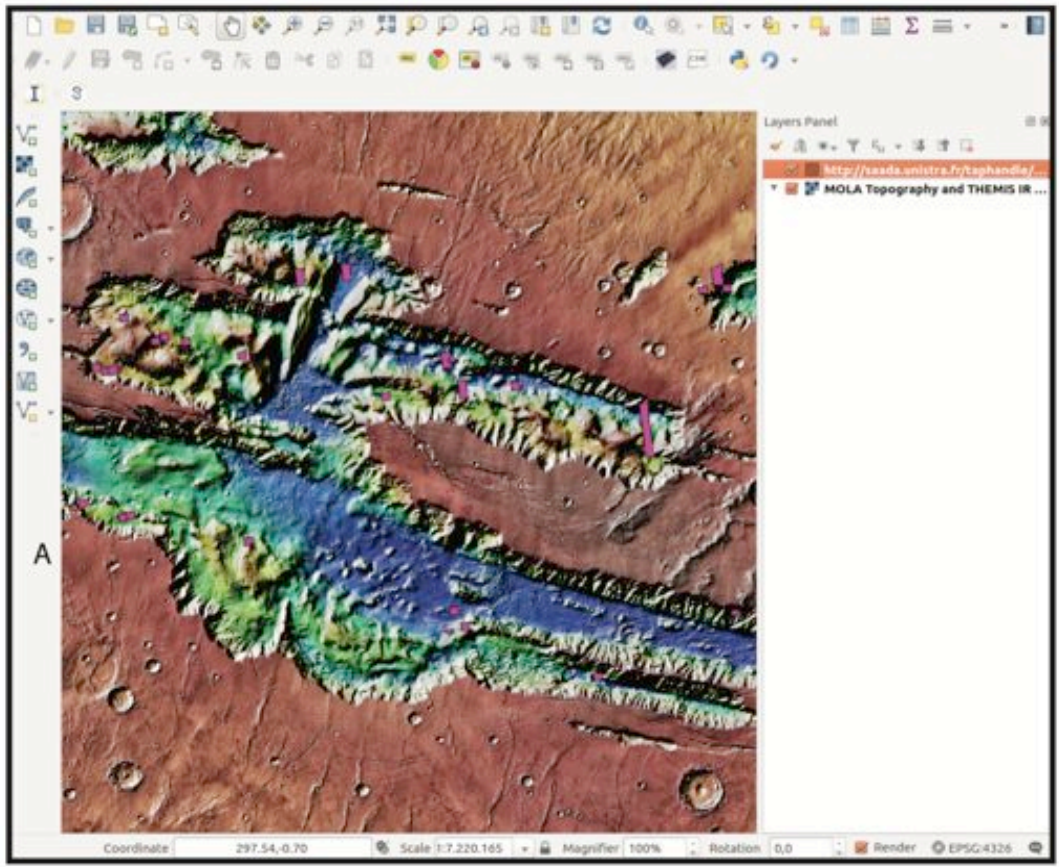

*Figure 10: Mars background map grabbed through WMS with CRISM footprints (in pink) served by an OGC coverage service, displayed in QGIS.*

Further connectivity is also planned with the **MarsSI** environment (Lozac'h et al., 2015; Quantin-Nataf et al., 2017). An automated procedure to set up an EPN-TAP service from existing planetary OGC Web Services is being devised, to populate the EPN-TAP parameters from existing OGC service functionalities. The compatibility between EPN-TAP and OGC standards is expected to help include non-surface data in GIS applications, such as atmospheric profiles and layers on Mars, or magnetic field line footprint locations on Jupiter satellites.

In addition, an extension of the FITS format was defined to include GIS-style coordinate reference systems and add georeferencing to FITS images (Marmo et al., 2016). This extension will be submitted to the IAU commission in charge of the FITS standard. Metadata conversion will be implemented using the *GDAL* library, making FITS format, one of the reference formats in VO applications, simply usable in an OGC context (Fig. 11). A possible further improvement is to study whether the *HiPS* multi-resolution tessellation system (Fernique et al., 2015) implemented in Aladin (used in Fig. 3b) can be adapted to planetary mapping on ellipsoids.

Finally, a coordinate conversion service may be provided both to support query writing in the VESPA user interface (at the same level as the name resolver and observatory list in Fig. 1) and to convert data retrieved from older data services. **TREPS** and SPICE's **WebGeoCalc** are being tested for this purpose, in association with a list of reference frames in the Solar System; GDAL functionalities are also being studied in this context.

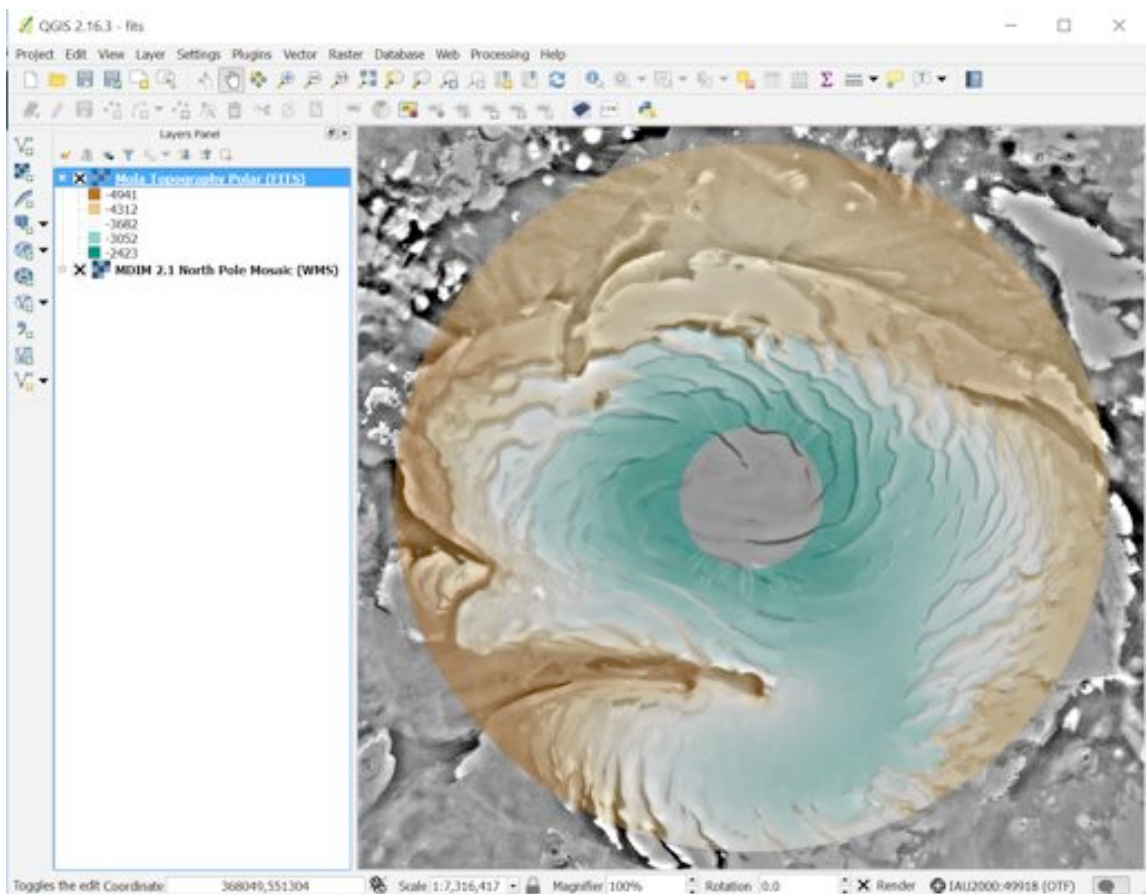

*Figure 11: WMS map of Mars' north pole and ice cap displayed in QGIS (Viking-based MDIM v2.1, credits: NASA/USGS). The superimposed colorized layer provides the MOLA digital elevation model, loaded from a FITS formatted file with geospatial keywords (browns low and blues high).*

### 3.3.1 Use case: revision of Mars crater database

A first example of using VO techniques to handle surface data is related to a database of Martian craters. Robbins' crater database (Robbins & Hynek, 2012) has been partly implemented as an EPN-TAP service at Jacobs University, including location and size provided as footprints displayable in Aladin (Fig. 3b). The database contains the morphological description of more than 384 000 craters larger than 1 km in diameter. However, this catalog is known to include false detections, and VO developments are the ideal framework for a catalog revision. A collaborative effort based on visual inspection of footprints on high-resolution images and the reference MOLA Digital Terrain Model, led by GEOPS in Orsay, will be finalized in 2017 (Lagain et al, in preparation). An inspection tool, Planetary Cesium Viewer, was developed at GEOPS for this purpose based on the Javascript Cesium framework (Fig. 12). The updated catalog will be distributed as an EPN-TAP service when ready.

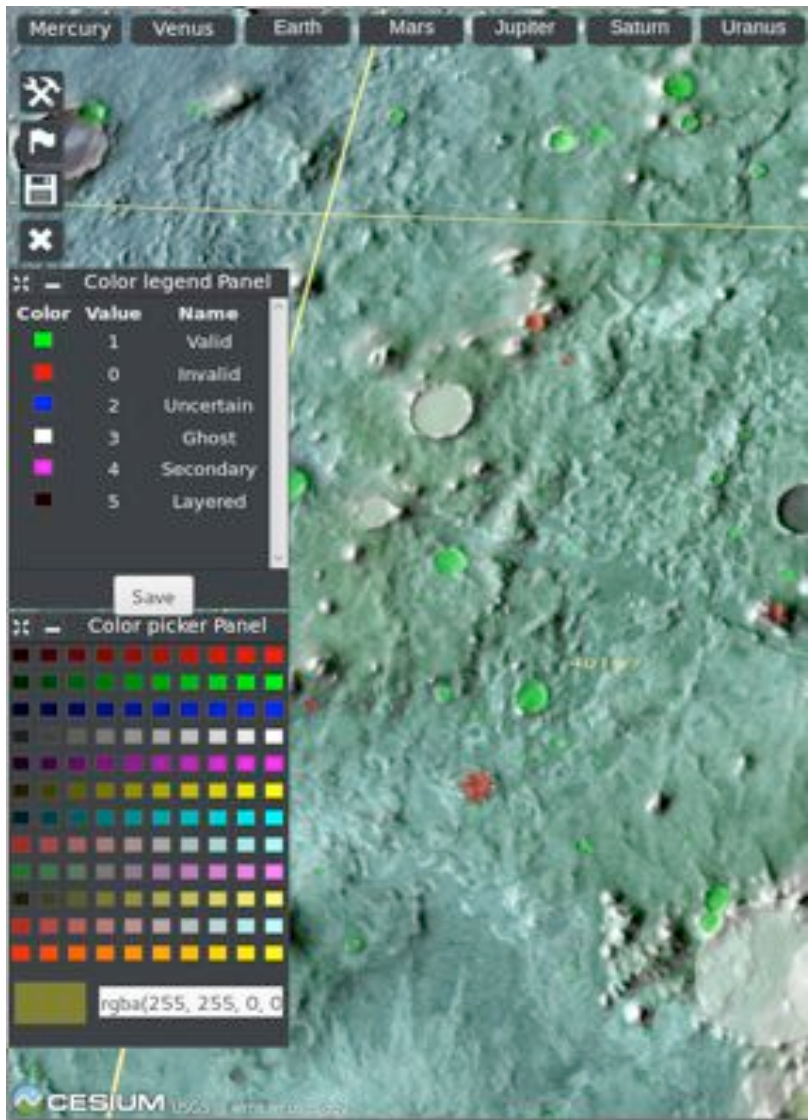

*Figure 12: the Planetary Cesium Viewer used for Robbins crater revision.*
*Red objects are examples of false detections in the current catalog, and are being corrected.*

### 3.3.2 Use case: Rosetta surface maps

Another use case is related to the analysis of surface changes and variability on comet 67P from VIRTIS/Rosetta observations (Rousseau et al., 2015). The original data are spectral cubes where each spectrum is localized precisely at the surface (the field of view is projected as a quadrilateral in the standard frame of 67P using the SPICE library, and viewing geometry is also computed). Data are read in IDL, which is also used to select observations of the nucleus and illumination conditions (e. g., maximum incidence angle), and to compute spectral parameters such as band depth, spectral slopes, single-scattering albedo retrieved from photometric fits, etc. These parameters are stored as various columns in a structure with one line / spectrum. Extra columns provide the location of the pixel center at the surface, and illumination conditions. The structure is written as a VOTable using the SSW IDL library. One such VOTable is produced for each period of one month, following the operation scheme of the Rosetta mission.

The VOTables are then ingested in TOPCAT. Relationships between parameters are easily plotted as simple graphs or density plots, and subgroups of spectra can be defined interactively from clusters in such plots. Cylindrical maps are also displayed with minimal efforts, using the coordinates provided with the dataset. Although the (varying) pixel size is not taken into account, TOPCAT performs a convolution on the fly when changing the size and shape of the symbol used as marker, so that overlapping data are summed in the plots to build actual density coverage maps (Fig. 13). 3D mapping is also available, but only on the sphere enclosing the shape model of the comet when using the longitude / latitude coordinates; however, the same display mode used with the 3D Cartesian coordinates provided in the dataset will map the data on the shape model itself, and will overcome the usual degeneration of the latitude/longitude frame on a concave body such as 67P.

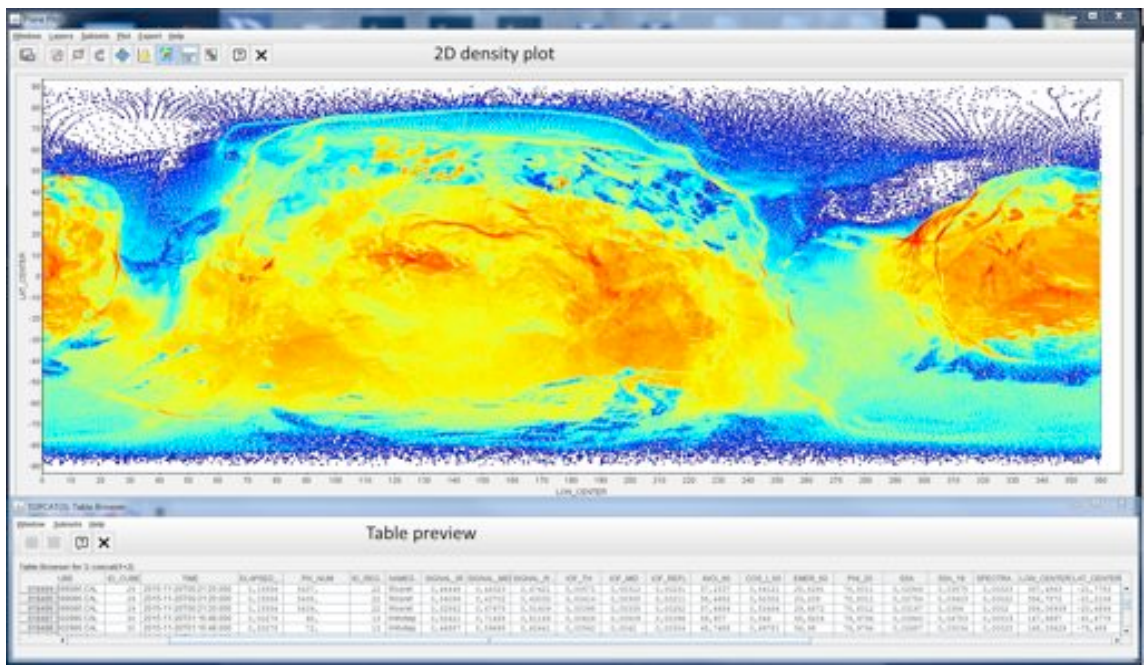

*Figure 13: VIRTIS/Rosetta coverage of comet 67P on a cylindrical map in TOPCAT. Observations performed with the mapping channel during two periods of 4 months (more than 3 millions spectra) are gathered here. The density plot outlines the two lobes (in red), whereas the coverage is much sparser in the neck region (in blue), which is more difficult to observe.*

Furthermore, TOPCAT includes the very powerful cross-correlation functions between tables, e. g., to associate spectra closer than a critical distance, or based on similar spectral properties. When applied to successive one-month periods, this allows the user to easily trace the seasonal evolution of particular spots at the surface, or to associate regions with similar behavior and to spot outliers. The resulting merged VOTables can be saved and re-ingested later on in IDL for further processing.

A variation on this use case is to first select in the VESPA search interface individual spectra at given locations and with predefined illumination conditions along the orbital phase of Rosetta. This will help, e. g., tracing the evolution of identified ice patches at the surface. The search is performed on a Rosetta data service similar to the extended VIRTIS/Venus-Express one mentioned above, which is currently installed in the team's institutes but not yet public. This demonstrates that private data services can be installed and interfaced to perform specific studies within a collaborative team, which may be especially useful during acquisition and proprietary periods, but also during the design phase of new instruments.

### 3.4 Magnetospheres and plasmas

Space plasma sciences have been developing interoperable tools during the last decade. Several projects worldwide are proposing efficient data sharing and display tools, such as: NASA's *SPDF* (USA) with its **CDAWeb** tool; the CDPP (France) with its **AMDA** tool; or *IUGONET* (Japan). The standards used to ensure this interoperability are coordinated by the *SPASE* group, which is proposing a metadata standard, as well as a registry. However, the developments have mostly been done with the Sun-Earth interaction science in mind. The goal of VESPA in this field is to bridge with planetary plasma sciences, as well as to enable efficient interfaces with neighboring science fields, e. g., with planetary atmospheres to study planetary aurora.

EPN-TAP data services open or in test phase in this field are listed in Table 1. New data services developed in the course of Europlanet 2020 are related to a modeling system ("Coupled Giant Planet Systems" at UCL), to a specific demonstrator of workflow for data analysis (in IAP, Prague), and to a large service of radio observations in support to the JUNO mission at Jupiter. One of the challenges in this science theme is to interface these new services with existing ones, e. g., AMDA and IMPEx.

In the first instance, the 'Coupled Giant Planet Systems' service will provide the user with outputs related to the UCL Magnetodisc Model. This model produces self-consistent magnetic field structures and plasma distributions related to the rapidly rotating magnetospheres of Saturn and Jupiter (Achilleos et al., 2010). These types of model outputs will thus provide a valuable means of interpreting field and particle observations from presently active spacecraft such as Cassini, Juno and, in the long term, the JUpiter ICy moons Explorer (JUICE).

The demonstrator of workflow for data analysis will aim at multi-dimensional measurements of planetary electromagnetic fields. It will calculate characteristics of electromagnetic waves from in-situ spacecraft measurements. These characteristics are the key signatures of fundamental processes in the solar wind and planetary magnetospheres. The demonstrator will be based on the PRASSADCO (PRopagation Analysis of STAFF-SA Data with COherency tests) analysis tool, developed originally in the frame of ESA's Cluster mission (Cornilleau-Wehrlin et al., 2003). It implements a number of methods used to estimate polarization and propagation parameters, such as the degree of polarization, sense of elliptic polarization, axes of polarization ellipse, the wave vector direction, the Poynting vector, and the refractive index. These methods have been previously used and validated for analysis of data from various space missions, including the STAFF-SA instruments onboard the four Cluster spacecraft and RPWS on Cassini.

The JUNO-Ground-Radio team is grouping all major decametric radio instruments world-wide and provides ground-based observations supporting the NASA

JUNO mission (Cecconi et al., 2017). The goal of this team is to provide a continuous monitoring of decametric Jovian radio emissions while the Juno spacecraft is orbiting Jupiter. This monitoring is particularly important due to the intrinsic anisotropy of the low frequency radio emissions: continuous and multi-point observations are required to interpret them. The coordination of observations is managed with a tool developed at Paris Observatory. Once the observations are available, each observatory is publishing them on a VESPA compliant server, so that the data is available through VESPA search interfaces. Currently, the Iitate radio observatory data are regularly published. The Nançay decameter array service is also regularly updated. Two other data services should be available soon: the Kharkov UTR-2 radio telescope (Ukraine) and the LWA1 (New Mexico, USA). The RadioJOVE spectrograph user group is also participating and the dataset will be available soon and regularly updated.

Space plasma observations are usually in-situ measurements of environmental physical parameters (plasma density, magnetic field, particle distribution function…). The main measurement axis is therefore temporal and data products are often treated as time-series, or spectrograms when time-series of spectra are available. The *CDF* format is usual in this field. Three tools are currently used for planetary plasma activities:

• TOPCAT handles CDF and can play a central role in the interaction between tools and services.

• AMDA includes a tool to plot time-series and spectrograms, as well as a conditional search tool (in addition to its database, which is accessible through an EPN-TAP server). AMDA tools can plot external data files retrieved through its internal EPN-TAP client, and can handle CDF, CSV or VOTable formats. Alternatively, an external interface such as VESPA can send data to AMDA tools through SAMP to display them.

• **Autoplot** has recently included SAMP support and accepts CDF files sent this way from the VESPA main user interface. It also implements a *HAPI* interface to retrieve times series from compatible services. HAPI is a standard set up in this field to smoothly extract times series from data services, regardless of the organization of data into files.

### 3.5 Heliophysics

EPN-TAP services currently implemented in this science field include some catalogs of solar structures inherited from the European *HELIO* program (Pérez-Suárez et al., 2012), the preexisting BASS2000 and CLIMSO image monitoring services (regularly receiving new data), as well as new services (radio observations from Nançay and Iitate, Japan). Although Heliophysics is not a dedicated science theme in the VESPA activity, EPN-TAP appears to be a convenient search protocol for solar data archives and several other external services are being designed from existing archives, either in visible light or radio waves.

### 3.6 Exoplanets

This science theme is currently restricted to the Encyclopedia of Extrasolar Planets in Paris (Schneider et al., 2011), which is by far the most used data service in VESPA, with ~ 50 000 unique accesses per month (summing EPN-TAP and other types of access). Although the web site includes plotting facilities (among other extra content, such as an extensive bibliography and a list of conferences), relationships between parameters are more easily visualized in TOPCAT.

A major activity is to upgrade the service infrastructure to face the rapidly growing number of new detections (more than 3500 planets and 3600 candidates at the time of writing), new types of objects (isolated planets, multi-star systems, etc), and new observed properties (atmospheric species, physical conditions, spectral simulations, etc).

Another topic is to develop an interface with the DACE/PlanetS database in Geneva (Buchschacher et al., 2015), so that the two services can be queried simultaneously through the EPN-TAP protocol. Common use cases for the two services will be defined, and will probably require a specific extension of EPN-TAP.

### 3.7 Solid spectroscopy

VESPA actions in this field include on one hand the support to the evolution of the GhoSST database into a much larger database infrastructure called SSHADE; on the other hand a unified access to several existing databases of mineral spectra, in support to surface studies of planets, satellites, and small bodies.

**SSHADE** (Solid Spectroscopy Hosting Architecture of Databases and Expertise) is a network of 20 European contributors from 8 different countries (Schmitt et al., 2015). It will extend the existing **GhoSST** (Grenoble astropHysics and planetOlogy Solid Spectroscopy and Thermodynamics) database (Schmitt et al., 2012) to a large set of contributor databases in the field of solid spectroscopy, including major ones. The on-going implementation phase will be followed by a phase of data documentation and validation, to ensure consistency and data quality. The resulting service will not only help extend the spectral databases of ices, minerals and organic material, but will also make the state-of-the-art laboratory data readily available as references to interpret observations of planets and small bodies, in particular from spacecraft. SSHADE includes a dedicated environment with visualization and processing tools for specialists; the databases will not only include measured spectra, but also derived data such as band lists and optical constants, relying on the very complete Solid Spectroscopy Data Model defined for this service. These databases will also be accessible from the simpler EPN-TAP interface, which is intended to speed up comparison with observational data and to allow for mass processing in the VESPA environment.

Other, existing databases of mineral reflectance spectra have been regularly populated in the past 20 years and are routinely used to interpret observations of planetary surfaces, e. g., on Mars, the Moon, or small bodies. An important project is therefore to extend EPN-TAP to define parameters describing samples of interest in terms of mineralogical composition, origin, grains size, mixing, possible processing, etc, as well as measurement technique and physical quantity (reflectance, emissivity, transmission, optical constants, etc). The current design phase is based on several spectral libraries, in particular the Berlin Rosetta Spectral library (minerals and meteorites in reflectance) and the CRISM spectral library (Murchie et al., 2007) which is currently searchable from a web form at the PDS Geosciences Node (http://speclib.rsl.wustl.edu/). Another obvious candidate for an EPN-TAP service is the Berlin Emissivity Database at DLR (Maturilli et al., 2008) which is supported in other Europlanet activities. With such descriptions available, the spectral libraries will be readily accessible by spectral fitting tools, but also by planetary GIS environments. This EPN-TAP extension will also be used to query databases providing other properties of extraterrestrial samples, e.g. lunar samples, meteorites, dust particles collected in orbit, etc.

A basic spectral fitting tool currently exists on the M4ast service web site (spectra of asteroids), but is not implemented as a VO-based workflow. To go beyond the usual process of manually selecting samples and fitting spectra, both band extraction techniques and band lists of laboratory spectra must be available to support automated identification of spectral signatures. A method to extract bands position and width on observational data is being developed and will be tested in a VO context (Erard, 2013). This is currently implemented under IDL, but can be embedded in a VO workflow. The results will be transmitted via SAMP as a VOTable describing each detected band, for use in tools such as TOPCAT or CASSIS.

A more technical but important action in this science theme is to adapt astronomical standards and tools to Planetary Science needs, where many spectral observations are acquired in reflected light and on extended objects. Current IVOA data models do not include the descriptors (*UCD*s and *UTypes*) identifying the corresponding physical quantities (radiance, I/F ratio, many sorts of reflectance, albedos and emissivities, etc), nor the usual physical units in the field. More generally, the various existing VO tools do not seem to handle our spectral data similarly at present, probably because of incomplete requirements provided by the standards. CASSIS is the tool of choice to assess the consistency of standards extension, since its (very reactive) development team is involved in VESPA. Visualization tools dedicated to bidirectional reflectance measurements are developed in Bern in the frame of the PlanetS program, and will be interfaced with the SSHADE databases in particular, possibly using IVOA standards. Formatting rules are also required to store spectra made of several overlapping ranges, which are commonly acquired from multi-detector instruments, cross-dispersion spectrometers, etc.

### 3.7.1 Use case: Comparing spectral observations with experimental data

A simple use case consists in searching spectra of asteroid Vesta in M4ast and comparing them with basaltic meteorites measured in reflectance from spectral libraries (Fig. 14). The main issue is to search for samples in the experimental databases, and to retrieve spectra that are sufficiently documented for later processing. It is clear from this use case that spectral thumbnails are important to quickly identify data of interest in the VESPA interface.

Both GhoSST and SSHADE present a sophisticated interface that requires some experience. EPN-TAP access will considerably speed up the search of spectra for casual users of these services. Since they will bypass the graphical interfaces, EPN-TAP queries from the VESPA portal will readily provide answers from SSHADE together with all other spectral services. A significant issue however is how to provide a description of the samples that is valid for a wide range of materials, including minerals but also organics, meteorites, or mixtures. SSHADE uses a complex system in which all constituents are described individually with many keywords, which can hardly be used without the help of a dedicated interface. A less complex but similar problem arises with other spectral libraries: the PDS spectral library uses a description with 8 different levels (global type, class, subclass, group, species…), while the Berlin one uses only 4 such levels, some with multiple values. In practice, it is difficult for the user to identify which keyword is expected to contain which descriptor, and it appears that databases do not always use a fully consistent description. The simple solution currently under assessment in EPN-TAP is to concatenate all available descriptors in a single list where substrings related to composition will be searched, e. g. "phyllosilicate" or "meteorite" as well as "kaolinite" or "CV3". Many tests are on-going to converge towards a satisfying description common to all data services, but fully efficient implementations will require some level of reprocessing of the composition information. Conversely, a dedicated form may be required to address this level of complexity in the main search interface.

Another important issue is to retrieve a data file that is properly documented and easily readable in TOPCAT or CASSIS, i. e. including explicit mention of quantities and units. Most databases propose an output format in CSV that is actually readable, but lacks any reference to the sample or measurement technique. Considering the amount of information associated to each measurement, only a VOTables permit to store the metadata conveniently with the data while maintaining readability. VOTable output therefore appears a key feature to use experimental spectral libraries efficiently in a VO context.

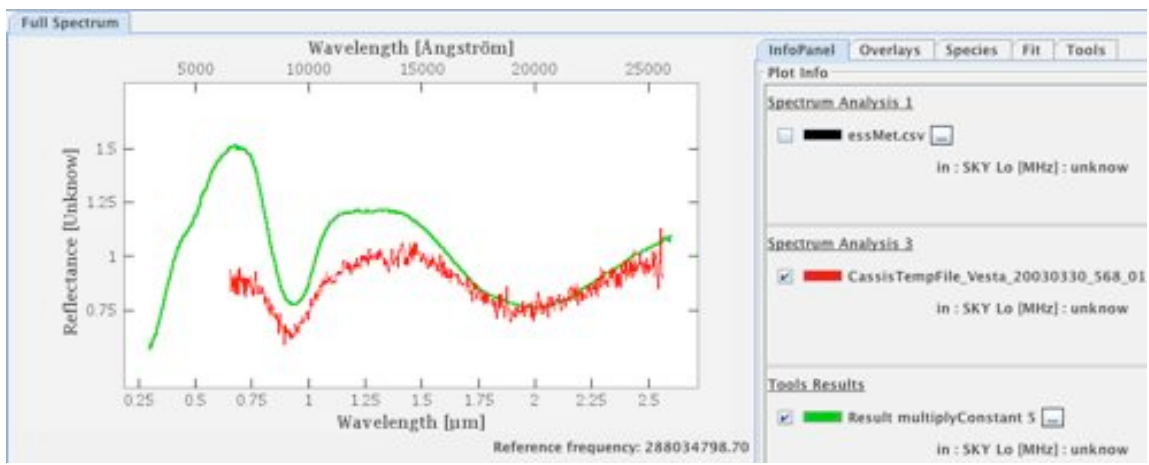

*Figure 14: Comparison between spectra of asteroid Vesta from the M4ast service and the ALH84001 SNC meteorite extracted from the PDS spectral library, in CASSIS. The two wide absorption bands are from pyroxenes present in basaltic materials.*

## 4- Prospects

VESPA currently offers a unique way to provide Planetary Science related databases with powerful search functions across all thematic fields. The program partners are developing new services in many fields of Solar System studies, and are making new data available with VO access. In addition, VESPA is connecting and customizing existing VO tools to support quick-look analysis, and makes more in-depth processing possible by identifying procedures and interfacing more familiar computing environments with the VO.

VESPA compliant data services are expected to multiply in the coming years, and will include archives of calibrated data on one hand, and derived data on the other hand. ESA is currently making the whole PSA content available in this system, and other space agencies are also considering VESPA as an additional way to make their data available and searchable. Several PDS nodes have demonstrated interest to implement a bridge toward (recent) PDS4 archives, which would extend EPN-TAP access to on-going and future space missions. This will provide the science users with a handy way to search large datasets and easily identify data of interest, based on observational or operational conditions.

Accessing derived data is of course even more profitable to the community. Although this can be coordinated at the level of a whole mission, typically by the Interdisciplinary Scientists (e. g., on Mars-Express or Cassini), research teams or individual researchers can also contribute and take advantage of the system directly by setting up a small data service to distribute results of analyses, e. g., from a published paper. All the VESPA documentation is freely available for this purpose, and the VESPA team will support such initiatives. During the course of the Europlanet program, VESPA opens a yearly call for new services; selected teams are invited to a 4-days implementation workshop where they acquire the know-how, install their own data server, and design their first service. This activity, together with the training sessions organized at EPSC and EGU conferences, is intended to build a community of both users and data providers.

In the long run, the VESPA infrastructure is expected to remain sustainable, because it only relies on professional standards. EPN-TAP data services are entirely supported by the TAP protocol from IVOA, therefore any TAP client can access their content, although missing the ability to query all services together. All developments are openly available on the VESPA GitHub repository, together with tutorials. The EPN-TAP protocol and its associated data model will be submitted to IVOA, as well as specific requirements related to Solar System studies (UCDs, Utypes, units, extensions of existing data models, etc). Other standards such as the Observatory and Spacecraft list and the Coordinate System list will be submitted to longer-lived consortia (IVOA, IPDA and IAU) depending on their scope. Finally, a Solar System Interest Group has just been approved by IVOA. It is expected to start from VESPA standards and contributions, and will build on them even after completion of the Europlanet program. VESPA will hopefully be a first step in a continuing process of making Solar System data archives consistent and interoperable, and will increase the science return of recent and future experiments.

## Acknowledgements

The Europlanet 2020 Research Infrastructure project has received funding from the European Union's Horizon 2020 research and innovation programme under grant agreement No 654208.

Additional funding was provided in France by the Action Spécifique Observatoire Virtuel, Programme National de Planétologie / CNRS-INSU, and Paris Astronomical Data Centre (PADC).

The authors wish to thank Trent Hare from USGS for fixing Figure 11 and FITS displays in QGIS. They are also grateful to two anonymous reviewers who greatly helped finalizing this paper by providing constructive comments.

## Annex 1: Protocols, standards, and facilities related to VESPA

CDF             Common Data Format. A self-described data format currently used in plasma physics, from NASA

                http://cdf.gsfc.nasa.gov/

Cone Search     An IVOA protocol to look around a fixed position in the sky

                http://www.ivoa.net/documents/latest/ConeSearch.html

DARTS           (Data Archive and Transmission System) JAXA's space missions data archive

                http://www.darts.isas.ac.jp/

DataLink        An IVOA standard to associate data and metadata spread on multiple files

                http://www.ivoa.net/documents/DataLink/20150617/index.html

EPN-TAP         Specific protocol to access Planetary Science data in Europlanet, currently in version 2 (Erard et al., 2014b)

FITS            (Flexible Image Transport System) One of the basic data formats in Astronomy, and a standard of IAU used by the IVOA (Pence et al., 2010)

                http://fits.gsfc.nasa.gov/fits_home.html

GDAL            (Geospatial Data Abstraction Library) Translator library for raster and vector geospatial data formats

                http://gdal.org/

GIS             Geographic Information Systems

HAPI            (Heliophysics Application Programmer's Interface) A protocol to retrieve times series from data services independently from data organization (John Hopkins Univ / APL) (Vandegriff et al., 2016)

                http://spase-group.org/docs/hapi-server

HELIO           HELiophysics Integrated Observatory, a FP7 EC-funded program

                http://www.helio-vo.eu/

HiPS            (Hierarchical Progressive Survey) An IVOA standard based on multi-resolution tessellation of the sphere providing the basis for visualizing data in a progressive way (Fernique et al., 2015)

                http://www.ivoa.net/documents/HiPS/20161122/index.html

IAU             International Astronomical Union, in charge of various standards in Astronomy (FITS format, Solar System nomenclatures and coordinate systems, etc)

                www.iau.org/

IPDA            (International Planetary Data Alliance) Reunion of space agencies for preservation and access to Planetary Science space archives.

                https://planetarydata.org/

IMPEx           Integrated Medium for Planetary Exploration, a FP7 EC-funded program

                http://impex-fp7.oeaw.ac.at

IUGONET         (Inter-university Upper atmosphere Global Observation NETwork) A Japan based project to make Earth's upper atmosphere databases interactive

http://www.iugonet.org/en/index.html

IVOA            International Virtual Observatory Alliance, in charge of the VO standards

http://www.ivoa.net/

LineList        A VAMDC protocol to access individual absorption lines, implemented in Specview (a variation on the older Spectral Line Access Protocol from IVOA)

MPC             (Minor Planet Center) An IAU-labelled and NASA-funded data center gathering and distributing parameters of small bodies of the Solar System. Their main database of orbital elements is called MCPOrb.

http://www.minorplanetcenter.net/

OGC             (Open Geospatial Consortium) Defines GIS standards used in commercial and open source applications

http://www.opengeospatial.org/

OpenPlanetary   A bottom-up initiative to address the needs of the Planetary Science community for data analysis problems (Manaud et al., 2016b)

http://openplanetary.co/about/

PDAP            (Planetary Data Access Protocol) Protocol to access planetary data space archives, developed and maintained by IPDA (latest version dated 16/4/2013)

https://planetarydata.org/projects/previous-projects/copy_of_2011-2012-projects/PDAP%20Core%20Specification%20/pdap-v1-0-16-04-2013/

PDS             (Planetary Data System) NASA's archiving system for Planetary Science data. Defines the de facto standard for Planetary Science space borne archives

https://pds.jpl.nasa.gov/

PDS4            Latest PDS archiving standard, only implemented on recent archives (Hughes et al., 2017)

http://pds.nasa.gov/pds4/about/what.shtml

PSA             (Planetary Science Archive) ESA's archive for Planetary Science data

http://www.cosmos.esa.int/web/psa/psa-introduction

SAMP            (Simple Application Messaging Protocol) IOVA protocol to exchange data between software applications (Taylor et al., 2012). Data are associated with MTypes (message-type) indicating a datatype ("table", "image", etc) and an action ("load", "highlight", etc). MTypes can be extended by developer agreement, so that uncommon types of data are readily supported. Such messages are routed to a common hub, and are received by all applications connected to the hub that subscribe to certain MTypes. All tools supporting SAMP are therefore automatically connected to the VESPA portal once they are launched. This makes it easy to dispatch data and actions to dedicated applications; for example Aladin will only receive images and tables of objects, a line selected from a table in TOPCAT will be highlighted in Aladin, etc.

http://www.ivoa.net/documents/SAMP/

SPASE           (Space Physics Archive Search and Extract) Defines data access standards for Space Physics

http://www.spase-group.org/

SPDF            (Space Physics Data Facility) A facility supporting data from most NASA Heliophysics missions at Goddard Space Flight Center

http://spdf.gsfc.nasa.gov

SPICE           Widely used software library for spacecraft geometry computations, from JPL (Acton et al., 2017)

http://naif.jpl.nasa.gov/naif/

| | |
|---|---|
| SSA | Simple Spectral Access protocol from IVOA |
| | http://www.ivoa.net/documents/SSA/20120210/index.html |
| TAP | (Table Access Protocol) An IVOA protocol to access astronomical catalogs (Nandrekar-Heinis et al., 2014) |
| | http://www.ivoa.net/documents/TAP/ |
| UCD | (Unified Content Descriptor) Controlled vocabulary used to broadly define physical quantities in the IVOA standards |
| | http://www.ivoa.net/documents/latest/UCD.html |
| UType | Description of data properties in IVAO standards, in relation with a Data Model – this is more specific than UCDs |
| | http://www.ivoa.net/documents/Notes/UTypesUsage/20130213/index.html |
| VAMDC | (Virtual Atomic and Molecular Data Center) A FP7 EC-funded program connecting many databases in the field (Dubernet et al., 2016) |
| | http://www.vamdc.eu/ |
| VESPA | (Virtual European Solar and Planetary Access) Planetary Science VO project in the Europlanet2020 program. Documentation is available on the web site |
| | http://www.europlanet-vespa.eu |
| | All developments in the frame of VESPA are available on a GitHub repository |
| | https://github.com/epn-vespa |
| | Discussions of VESPA related matters, including service implementation procedures, are available on the project wiki: |
| | https://voparis-confluence.obspm.fr/ |
| | VESPA is also the generic name of the EPN-TAP/PDAP client and main user interface: http://vespa.obspm.fr |
| VizieR | A database with VO access grouping thousands of astronomical catalogues and data associated to publications at CDS, Strasbourg (Ochsenbein et al., 2000) |
| VOTable | Self-described XML data format. One of the basic data formats in the VO |
| | http://www.ivoa.net/Documents/VOTable/ |
| WCS | (Web Coverage Service) An OGC protocol to access footprints of maps and images |
| WCS | (World Coordinate System) This acronym also refers to the coordinate system associated to the FITS format (Greisen & Calabretta, 2002) |
| WFS | (Web Feature Service) An OGC protocol to access vector data |
| WMS | (Web Map Service) An OGC protocol to access maps |

## Annex 2: Tools of interest in VESPA context

Tools benefiting from direct VESPA/Europlanet 2020 support or developed in the frame of VESPA are underlined.

| | |
|---|---|
| 3Dview | 3D viewer for planetary objects, spacecraft trajectories and data developed by GFI/Toulouse and CDPP/IRAP/Toulouse (initiated under CNES contract) (Génot et al., 2017a) |
| | http://3dview.cdpp.eu/ |
| Aladin | Image plotting tool and sky atlas, developed by CDS/Strasbourg (Bonnarel et al., 2000) |

| APERICubes | PDS spectral cubes on-line reader & viewer, currently supporting VIRTIS/Venus-Express data (Savalle et al., 2016) |

http://voparis-apericubes.obspm.fr/apericubes/js9/demo.php

| AMDA | Automated Multi-Dataset Analysis, at CDPP/Toulouse. Environment including databases and display tools for plasmas (Jacquey et al., 2010) |

http://amda.cdpp.eu

| AutoPlot | Interactive plotting tool for space physics data and time series (University of Iowa) (Faden et al., 2010) |

http://autoplot.org

| CASSIS | Spectral analysis tool with VO interface, developed by IRAP/Toulouse (Vastel et al., 2015) |

http://cassis.irap.omp.eu/

| CDAWeb | (Coordinated Data Analysis Web) Web-based tool to access and plot planetary plasma data |

http://cdaweb.sci.gsfc.nasa.gov/index.html/

| Cosmographia | SPICE-based interactive tool used to produce 3D visualizations of planet ephemerides, spacecraft trajectories, instrument field-of-views and footprints, from JPL (Acton et al., 2017) |

http://naif.jpl.nasa.gov/naif/cosmographia.html

| EPN-TAP Google sheet add-on | EPN-TAP client developed at IWF/Graz |

https://drive.google.com/file/d/0B5hOkAv922Z6OU9fN3RyQ0JRZGc/view

| GDL | Open source clone of IDL, a powerful but proprietary computing environment routinely used in many fields of Planetary Science. GDL has significantly restricted graphic capacities but may be more rapid in processing (Coulais et al., 2012) |

http://gnudatalanguage.sourceforge.net/

| GhoSST | (Grenoble Astrophysics and Planetology Solid Spectroscopy and Thermodynamics) Solid-phase spectroscopy service including databases and tools, at IPAG/Grenoble (Schmitt et al., 2012) |

http://ghosst.osug.fr/

| MarsSI | Mars web-GIS and processing / data server application, a European Research Council-funded project at University of Lyon (Quantin-Nataf et al., 2017): |

https://emars.univ-lyon1.fr/MarsSI/

| Matrix of ground-based facilities | List of observatories with submission interface (open to amateurs) at IWF/Graz, inherited from Europlanet FP7 (Scherf et al., 2011) to be coordinated with the VESPA Observatory list (Cecconi et al., 2015a) |

http://iwf.oeaw.ac.at/matrix/

| MATISSE | 3D viewer for planetary objects, initiated under contract from ASI (Italian Space Agency) (Zinzi et al., 2016) |

https://tools.asdc.asi.it/matisse.jsp

| Miriade | VO-compliant ephemeris service with visibility charts (VISION) at IMCCE/Paris Observatory (Berthier et al., 2009) |

http://vo.imcce.fr/webservices/miriade/?forms

| Mizar | 3D web client for discovering and visualizing geospatial data on both sky and planets, from CNES |

Planetary Cesium Viewer      3D web client to visualize planetary geospatial data, based on the open source Cesium environment, developed by GEOPS/Orsay

http://134.158.75.177/

PlanetServer      Web-GIS focused on planetary surfaces (Jacobs University) (Oosthoek et al., 2014), (Marco Figuera et al., 2017)

http://planetserver.jacobs-university.de/

Propagation Tool      Interactive tool to track energetic particles in the heliosphere, at CDPP/IRAP, Toulouse (Rouillard et al., 2017; André et al., 2017)

http://propagationtool.cdpp.eu/

QGIS      Open source Geographic Information System

http://www.qgis.org/

Specview      Spectral analysis tool with access to reference data (STScI) (Busko, 2002)

http://www.stsci.edu/institute/software_hardware/specview

SPLAT-VO      (SPectraL Analysis Tool) VO graphical tool for displaying, comparing, modifying and analysing astronomical spectra (developed by Durham University, maintained by Heidelberg University)

http://www.g-vo.org/pmwiki/About/SPLAT

SSHADE      (Solid Spectroscopy Hosting Architecture of Databases and Expertise) Multi-contributor spectral database infrastructure in IPAG/Grenoble, expanding GhoSST (Schmitt et al., 2015)

https://blog.sshade.eu/

SsODNet      Solar System Open Database Network at IMCCE, Paris Observatory. Currently used as a name resolver for Solar System objects in VESPA (Berthier et al., 2007)

http://vo.imcce.fr/webservices/ssodnet/?forms

SSW      An IDL library for Solar Physics which includes VO access routines

http://www.lmsal.com/solarsoft/

TapHandle      Search interface for individual TAP services (Univ. Strasbourg) (Michel et al., 2014)

http://saada.unistra.fr/taphandle

TAP shell      Command line interface to query TAP servers (Heidelberg University)

http://vo.ari.uni-heidelberg.de/soft/tapsh

TOPCAT      (Tool for OPerations on Catalogues And Tables) VO analysis tool for tabular data (Bristol University) (Taylor, 2011)

http://www.star.bris.ac.uk/~mbt/topcat

TREPS      (Transformation de REpères en Physique Spatiale) Tool devoted to coordinate transformations, at CDPP (Génot et al., 2017b)

http://treps.cdpp.eu/

VESPA      is also the generic name of the EPN-TAP/PDAP client and main user interface: http://vespa.obspm.fr

VOSpec      VO spectral analysis tool from ESA/ESAC (ESAVO Team, 2012)

http://www.cosmos.esa.int/web/esdc

WebGeoCalc      Web-based graphical user interface to SPICE geometry computations, from JPL (Acton et al., 2017)

https://naif.jpl.nasa.gov/naif/webgeocalc.html

Table 1: Data services using the EPN-TAP protocol at the date of writing. Test implementations currently have private access only.

| Service name / Field | Description | EPN-TAP service location | Status |
|---|---|---|---|
| **Small bodies** | | | |
| M4ast | Visible-NIR CCD spectroscopy of asteroids, Popescu et al. (2012) | Paris Observatory, IMCCE | Open |
| TNOsarecool | Results from Spitzer and Herschel and compilation of published TNO properties, Lellouch et al. (2013) | Paris Observatory, LESIA | Open |
| IKS/Vega-1 | IR spectroscopy of 1P/Halley, Combes et al. (1988) | Paris Observatory, LESIA / PDS small bodies | Open |
| BaseCom | 30 years of radio observations of comets from Nançay, Crovisier et al. (2015) | Paris Observatory, LESIA | Open |
| Illu67P | Illumination configurations on 67P/Churyumov-Gerasimenko, Beth et al. (2017) | IRAP, Toulouse | Open |
| DynAstVO | DynAstVO, computation of orbital parameters of NEOs daily updated from new observations, Desmars et al. (2016) | Paris Observatory, IMCCE | Open |
| MPC | VO implementation of asteroid catalog from IAU/Minor Planets Center, Rudenko (2016) | Heidelberg Univ. / Paris Observatory | Open |
| **Atmospheres** | | | |
| Titan CIRS | Vertical profiles by CIRS/Cassini, Vinatier et al. (2010), Vinatier et al. (2015) | | Open |
| SOIR | Vertical profiles of Venus by SOIR/Venus-Express, Trompet et al. (2017) | IASB/BIRA, Brussels | Open |
| SPICAM | Vertical profiles of Mars by SPICAM/Mars-Express, Forget et al. (2009), Lebonnois et al. (2006), Määttänen et al. (2013) | LATMOS, Paris | Open, being expanded |
| MCD | Vertical profiles extracted from the Mars Climate Database, Lewis et al. (1999) | LMD, Paris | Open |
| VVEx | VIRTIS/Venus-Express calibrated dataset | Paris Observatory, LESIA | Open |
| abs_cs | Absorption cross-sections of gaseous species | IAA-CAB-INTA-CSIC, Granada | Open, being populated |
| **Surfaces** | | | |
| USGS_WMS | Planetary maps from USGS Astrogeology, Hare et al. (2017) | Jacobs University, Bremen | Open |
| CRISM | CRISM/MRO calibrated spectral cubes on Mars | Jacobs University, Bremen | Open |
| Mars craters | Robbins' crater database, Robbins | Jacobs University, | Open |

| | and Hynek (2012) | Bremen | Test implementation |
| | Revised version of Robbins' database | Orsay, GEOPS | |

**Magnetospheres**

| | | | |
|---|---|---|---|
| AMDA | CDPP plasma database, Jacquey et al. (2010) | IRAP, Toulouse | Open |
| APIS | Auroral Planetary Imaging and Spectroscopy, Lamy et al. (2015) | Paris Observatory, LESIA | Open |
| MAG | MAG/Venus-Express dataset | IWF, Graz | Open |
| RadioJOVE | Amateur radio measurements of Jupiter | Paris Observatory, LESIA | Open, being populated |
| IMPEx | Selected simulation databases from the IMPEx program | IWF, Graz | Open |
| NDA | Radio monitoring of Jupiter and the Sun from the Nançay Decameter Array | Paris Observatory, LESIA | Open, being populated |
| Iitate | Decametric observations of Jupiter from Iitate observatory | Tohoku University, Japan | Open, being populated |
| UTR-2 | Kharkov UTR-2 radio telescope | NASU/RINANU, Ukraine | Test implementation |
| Coupled Giant Planet Systems | UCL Magnetodisc Models, Saturn & Jupiter, Achilleos et al. (2010) | UCL, London | Test implementation |

**Heliophysics**

| | | | |
|---|---|---|---|
| HFC1AR | HELIO catalogue of active regions | Paris Observatory, LESIA | Open |
| HFC1T3 | HELIO catalogue of type 3 radio bursts | Paris Observatory, LESIA | Open |
| CLIMSO | Image monitoring of the Sun from Pic du Midi, Koechlin (2012) | IRAP, Toulouse | Open, receiving new data |
| BASS2000 | Ground-based solar survey archive | Paris Observatory, LESIA | Test implementation |
| PPARC_R | Decametric observations of the Sun from Iitate observatory | Tohoku University, Japan | Open, being populated |
| RadioSolarDatabase | Radio monitoring of the Sun from the Nançay Decameter Array | Paris Observatory, Nançay | Test implementation |

**Exoplanets**

| | | | |
|---|---|---|---|
| Encyclopedia of Extrasolar Planets | Catalogue of Extrasolar Planets, Schneider et al. (2011) | Paris Observatory, LUTh/LESIA | Open |

**Solid spectroscopy & samples**

| | | | |
|---|---|---|---|
| SSHADE | Contributive ice & mineral spectroscopy databases, Schmitt et al. (2015) | IPAG, Grenoble | Test implementation |
| PDS spectral library | Contributive mineral & rock spectroscopy database, Murchie et al. (2007) | Paris Observatory, LESIA / PDS Geosciences | Test implementation |
| Berlin Rosetta Spectral Library | Contributive mineral & rock spectroscopy database | DLR, Berlin | Test implementation |
| NASA dust | NASA's Cosmic dust catalogs 15 | IAPS/INAF, Rome | Open, being |

| catalogue | and 18 | | redesigned |
|---|---|---|---|
| **Generic services** | | | |
| PVOL | Amateur images of planets, Hueso et al. (2017) | UPV/EHU, Bilbao | Open, accepting submissions |
| BDIP | Historical planetary images from an IAU program, Drossart et al. (2002) | Paris Observatory, LESIA | Open |
| Planets | Physical properties of planets & satellites, reference data | Paris Observatory, LESIA | Open, being expanded |
| Observatories & Spacecraft list | Build from IAU observatory list and other sources; interfaced with the Matrix of ground-based facilities and name resolver | Paris Observatory, LESIA and IWF, Graz | Test implementation |
| Selene | Datasets from JAXA's lunar mission Selene | Paris Observatory, LESIA | Test implementation |
| PSA | ESA's Planetary Science Archive | ESAC, Madrid | Test implementation |